\documentclass[12pt,english]{article}
\usepackage{amsmath}
\usepackage{amsthm}
\usepackage{amssymb}
\usepackage[authoryear]{natbib}
\usepackage[unicode=true,
 bookmarks=true,bookmarksnumbered=false,bookmarksopen=false,
 breaklinks=false,pdfborder={0 0 1},backref=false,colorlinks=false,hyperfootnotes=false]{hyperref}
\usepackage{graphicx}

\usepackage{fullpage}
\usepackage{setspace}
\usepackage[bottom, flushmargin]{footmisc}
 
\theoremstyle{definition}
\newtheorem{defn}{\protect\definitionname}
\theoremstyle{plain}
\newtheorem{lem}{\protect\lemmaname}
\providecommand{\definitionname}{Definition}
\providecommand{\lemmaname}{Lemma}
\theoremstyle{plain}
\newtheorem{cor}{Corollary}

\theoremstyle{plain}
\newtheorem{proposition}{Proposition}

\theoremstyle{plain}
\newtheorem{assum}{Assumption}

\begin{document}

\title{Optimal dynamic insurance contracts\thanks{
    This paper supersedes an earlier paper entitled ``Dynamic Competitive Insurance''.
} 
\thanks{
    I would like to thank Dirk Bergemann, Johannes Hörner and Larry Samuelson
    for encouragement and feedback on this project. I would also like
    to thank Brian Baisa, Eduardo Souza Rodrigues, Jaqueline
    Oliveira, Bruno Badia, Sofia Moroni, Marina Carvalho, Marcelo Sant'anna,
    Sabyasachi Das, Eduardo Faingold, Li Hao, Mike Peters, Alex Frankel and the seminar participants at the 
    Yale Micro lunch and Summer workshop,
    2022 EUI Alumni Conference,
    NYU,
    Olin Business School,
    University of British Columbia,
    Columbia Business School,
    Notre Dame,
    Boston University, 
    Warwick, 
    EPGE/FGV,
    EESP/FGV and the
    Catholic University of Rio de Janeiro
    for helpful comments.
}}

\author{Vitor Farinha Luz\thanks{
        University of British Columbia. Email: \href{mailto:vitor.farinhaluz@ubc.ca}{vitor.farinhaluz@ubc.ca}
    }
}

\maketitle

\begin{abstract}
    I analyze long-term contracting in insurance markets with asymmetric information.
    The buyer privately observes her risk type, which evolves stochastically over time.
    A long-term contract specifies a menu of insurance policies, 
    contingent on the history of type reports and contractable accident information.
    The optimal contract offers the consumer in each period a choice between 
    a perpetual complete coverage policy with fixed premium
    and a risky continuation contract in which current period's accidents may affect not only within-period consumption (partial coverage) but also future policies.

    The model allows for arbitrary restrictions to the extent to which firms can use accident information in pricing.
    In the absence of pricing restrictions, accidents as well as choices of partial coverage are used in the efficient provision of incentives.
    If firms are unable to use accident history, 
    longer periods of partial coverage choices are rewarded, 
    leading to menus with cheaper full-coverage options and more attractive partial-coverage options; and allocative inefficiency decreases along all histories.
    
    These results are used to study a model of perfect competition, where the equilibrium is unique whenever it exists, as well as the monopoly problem, where necessary and sufficient conditions for the presence of information rents are given.
    
\end{abstract}

\section{Introduction}


A vast majority of insurance contracts cover risks that are present over many periods, such as auto, auto or home insurance.
This allows insurance firms to benefit from the use of dynamic pricing schemes, where consumers premium and coverage evolve over time and incorporate any information observed over the course of their interaction.
The dynamics of coverage and premium are a central issue in several insurance markets including
auto insurance (\cite{dionne1994adverse}, \cite{cohen2005asymmetric}),
health insurance (\cite{handel2015equilibria}, \cite{atal2019lock}, \cite{atal2020long})
and life insurance (\cite{HendelLizzeri2003}, \cite{daily2008does}).

This observation raises the fundamental theoretical question of 
what dynamic pricing schemes arise as a result of profit maximization and what are their consequences for coverage and premium dynamics.
The theoretical insurance literature has focused on scenarios with either symmetric risk information\footnote{
    See \cite{HendelLizzeri2003} and \cite{ghili2019welfare}. 
} 
or permanent risk types.\footnote{
    See \citet{cooper1987multi} and \citet{dionne1994adverse}.
}
In this paper, I characterize profit maximizing long-term contracts in repeated interactions with evolving, persistent and private risk information, and apply this characterization to study equilibrium outcomes under perfect competition and monopoly environments.


In my model, a risk-averse consumer (she) may incur incidental losses in each period, such as a car accident (auto insurance) or expenditures from medical procedures (health insurance); 
and the probability distribution over losses is determined by her risk-type, which is privately known by the consumer and follows a persistent Markov process.
A high (low) type has a higher (lower) expected expected income, net of losses, in each period.

I assume that an insurance firm offers a long-term contract to the consumer after she has observed her initial risk-type.
A long-term contract represents a commitment to a schedule, in which a menu of insurance policies (referred to as flow contracts) with different premiums and coverage levels is offered to the consumer in each period. 
Crucially, the menu offered on a given period may depend on previous choices made by the consumer as well as any available information regarding accident history.
Motivated by common regulatory policies that limit the use of accident history in pricing (\cite{handel2015equilibria}, \cite{farinha2021risk}),
I allow for restrictions on the amount of information about past accidents that can be explicitly used by firms in setting insurance policy offers to be made to the consumer.
These restrictions include as special cases both fully contingent contracts, where the whole history of accidents can be used, and realization-independent contracts, where the firm cannot use any past accident information in pricing.

The first part of the paper (Sections \ref{Sec:profit-maximizing-contracts}-\ref{sec:Utility-Distortion-dynamics}) considers fixed discounted utility levels for the consumer, dependent on her initial risk-type, and characterizes the profit maximizing long-term contract that delivers these utility levels. 
This allows for the derivation of qualitative properties of optimal contracts that hold in both the competitive (Section \ref{sec:competitive-analysis}) and monopolistic (Section \ref{sec:monopoly}) settings studied in the second part of the paper, where each market setting corresponds to different equilibrium utility levels for the consumer.


I show that the optimal contract features a simple pricing scheme: in every period the consumer chooses between a complete coverage insurance policy with a perpetual fixed premium (efficient in this environment), and a partial coverage policy, in which case future offers will depend on additional accident realizations. 
The consumer is induced to choose the full coverage option when having a low-type realization, and to choose partial coverage when having successive high-type realizations.

In each period, an insurance firm has access to two sources of information: accident information --- which is directly observed by the firm --- and the consumer's choice (or announcement in a direct mechanism) which provides endogenous information about their type. 
In the optimal mechanism, policy offers use both pieces of information to efficiently screen different risk types. 
Accident information that is indicative of high-types is rewarded with higher continuation utility, in the form of more attractive future contracts. In a similar fashion, the choice of partial coverage in a given period is indicative of a high-type and is rewarded with higher continuation utility.


The characterization of distortion and coverage dynamics is challenging due to two technical challenges.
First, flow contracts are multi-dimensional objects, describing a premium and coverage for each possible loss level. 
Second, the presence of risk aversion leads to a non-separability issue absent in quasi-linear environments.
The separation of consumers with different types requires the introduction of distortions, in the form of partial coverage.
The marginal efficiency loss from the introduction of distortions potentially depends on the underlying utility level obtained by the consumers in a given period. 
As a consequence, the optimal contract requires a joint consideration of the issues of 
(i) intertemporal allocation of utility to be provided ot the consumer as well as 
(ii) the efficient spreading of distortions over time.

I tackle both of these issues by introducing an auxiliary static cost minimization problem in Section \ref{sec:auxiliary-problem},
which finds the optimal flow contract that (i) delivers a certain utility level if consumed by a high-type, and (ii) provides a fixed punishment if chosen by a low-type pretending to be high-type.
In Section \ref{sec:Utility-Distortion-dynamics} two optimality conditions are derived which characterize both the efficient spreading of 
utility and distortions over time, in terms of the solution to this auxiliary cost minimization problem.


Under a mild condition on the consumer's Bernoulli utility function,\footnote{
    This condition requires that the consumer's absolute risk aversion does not decrease ``too quickly'' with consumption, which is guaranteed if the consumer has non-decreasing absolute risk aversion.
} 
the auxiliary cost function is supermodular, i.e., the marginal cost of distortions in a flow contract increases in the utility level generated by it.
Under cost-supermodularity, the intertemporal optimality conditions can be used to obtain sharp characterization of the dynamics of utility and distortions.

For the case of realization-independent contracts, we show that distortions --- measured in terms of the consumer's exposure to risk --- are strictly decreasing over time. As a consequence, the optimal mechanism has decreasing distortions along \textit{all} paths.
When it comes to the flow utility dynamics, we show that consecutive high-type announcements --- which are revealed by the choice of partial coverage --- are rewarded by leading to the offer of partial contract offers that generate higher flow utility, as well as lower distortion.

For the case of fully contingent contracts, distortions and utility depend on the history of income realizations and hence are stochastic, even conditioning on types. 
I present results for two special cases.
First, I focus on the case of two periods and show that the firm has an incentive to reward high-type announcements and reduce distortions. 
I recast the firm's contract design problem as a cost minimization one and show that, 
when averaging over possible first-period income realizations, 
the expected marginal cost of distortions in flow contracts decrease over time, while the expected marginal cost of flow utility increases over time following the choice of partial coverage.
Second, 
we focus on the illuminating knife-edge case of constant relative risk aversion utility
$u(c) = \sqrt{c}$,
which corresponds to a separable auxiliary cost function,
and an arbitrary number of periods.
Consecutive choices of partial coverage --- which corresponds to high-type announcements ---
lead to a path of insurance policies that depend on the history of income realizations. I show that these policies have lower distortions and higher flow utility levels over time, when averaging over possible income realizations.


In Section \ref{sec:competitive-analysis}, we consider a competitive model in which multiple firms make long-term contract offers to a consumer who is able to commit to a long-term contract. I extend the characterization of \cite{rothschild1976equilibrium} to our environment and show that a unique outcome, featuring the flow utility and distortion dynamics aforementioned, can be sustained by a pure strategy equilibrium, and obtain necessary and sufficient conditions for such an equilibrium to exist.

The assumption of consumer commitment is reasonable in markets with high search or switching costs which preclude or inhibit the consumer's consideration of or transition to new firms.
In the absence of such frictions, the commitment assumption is potentially with loss.
If the low-type is an absorbing state of the types' Markov process (such as a chronic condition in health insurance), 
I propose a simple competitive model with consumer reentry in the market and show that firms' endogenous beliefs about the type of consumer searching for a new contract discourages consumer firm switching and, as a consequence, the commitment outcome can be sustained in a Perfect Bayesian Equilibrium of the no-commitment model.
This follows from the fact that, the optimal contract with commitment is back-loaded: both the partial- and the full-coverage policy options within the optimal menu become more attractive over time.


Section \ref{sec:monopoly} studies the case of monopoly, where the consumer's (type-dependent) outside option is given by zero coverage. 
The optimal contract is always separating and hence features the flow utility and distortion dynamics aforementioned. 
The consumer with initial high-type has no information rent, having total utility equal to that in the absence of insurance. 
I present a condition that is necessary and sufficient for the optimal contract to leave information rents to the consumer with initial low-type, meaning that her utility is strictly higher than the no-insurance option.

\subsection*{Related literature}\label{subsec:Related-literature}

This paper contributes to the literature on competitive screening, initiated by the seminal contributions of \citet{rothschild1976equilibrium} and \citet{wilson1977model},
which studies situations in which private information leads to inefficiencies in competitive markets.
\citet{rothschild1976equilibrium} considers insurance markets in which customers have private information regarding their risk characteristics.
They show that competition leads to a unique equilibrium in which high-type consumers, who have lower accident probabilities, are screened by the choice of partial insurance at better premium rates (per unit of coverage).

\citet{cooper1987multi} extends the analysis of \citet{rothschild1976equilibrium}
to a multi-period setting in which consumers have fixed risk types and
full commitment. 
They find that optimal long-term contracts use experience rating as an efficient sorting device in addition to partial coverage. 
The optimal contract features a single type announcement in the first period and
customers make no further choices. 
\cite{dionne1994adverse} allows for renegotiation and competition in the same fixed-types model and find equilibria with semi-pooling.

My analysis is also related to the dynamic mechanism design literature (see \citet{courty2000sequential}, \citet{battaglini2005long}, \citet{esHo2007optimal}, \citet{pavan2014dynamic}), which differs from the insurance model considered here in two relevant ways.
First, two sources of information exist in each period: the consumer's coverage choice (or type announcement) as well as the accident information that can be directly used in future offers.
This is in contrast with standard screening models in which the only source of information to the mechanism designer over the course of an interaction is the series of choices or announcements made by the consumer.
Second, the presence of curvature in preferences implies that the premium dynamics are completely pinned down in the optimal contract. 
This is in contrast with quasi-linear environments considered in the literature, where transfers are only pinned down up to their total ex-post present value.\footnote{
    This statement assumes equal discount rates for both involved parties. \cite{krasikov2021implications} 
    show that, if the mechanism designer is more patient than the consumer,
    the efficiency gain from front-loading completely pins down transfers in the optimal mechanism.
}

A notable exception on both fronts is \cite{garrett2015dynamic}, which studies the optimal compensation scheme for a manager who is privately informed about his persistent productivity type in two periods. Production in each period is observable and corresponds to the sum of productivity and unobservable effort.
Instead of studying a relaxed problem, their approach is to construct two sets of perturbations which retain implementability in the original mechanism design problem and exploit these perturbations to characterize the dynamics of distortions, which correspond to under-provision of effort, in the optimal mechanism. 
In their model, even the implementation of constant efforts (which only depend on first period productivity) requires the introduction of additional consumption variation in the second period, which is costly for a risk averse manager.
This is a crucial difference to my framework, where distortions \textit{correspond to} consumption variability within a period.
As a consequence, they show that, if the manager is risk neutral and types are sufficiently persistent, distortions in the optimal mechanism increase over time on average. The reverse is true if the manager is sufficiently close to risk neutrality.

Another closely related paper is \citet{battaglini2005long}, which considers the design of dynamic selling mechanisms in a monopoly setting where customer's valuation follows a two-state Markov chain. 
In the optimal mechanism, production becomes efficient when the customer obtains his first high valuation and converges to the efficient level along the history path of consecutive low valuations, which is analogous to our characterization of distortions for realization-independent contracts (Proposition \ref{prop:RI_monotonicity}). 
Even though the productive allocation is fully characterized,
the pricing scheme is not unique.

\cite{HendelLizzeri2003} study long-term life insurance contracts with symmetric information and one-side commitment. The optimal contract features front-loaded payments, 
which remedies the consumer's inability to commit by locking in consumers who expect to pay lower premiums later in the relationship. 
Using data from the U.S. life insurance market, they show that front-loading is a common feature of contracts, with more front-loading being negatively correlated with net present value of premiums.

\cite{ghili2019welfare} study the same contract design problem in health insurance markets. 
They characterize the optimal dynamic contract assuming symmetric risk information and one-sided commitment, which features full insurance in each period. 
The presence of one-sided commitment precludes complete consumption smoothing over time, hence the consumer is offered a \textit{consumption floor}, which is adjusted upwards whenever the consumer's outside option is attractive enough so that the participation constraint becomes binding. 
Finally, they estimate a stochastic Markovian model of health status and medical expenses using data from the state of Utah and numerically find the optimal long-term contract for the estimated model parameters.

The analysis of commitment in Subsection \ref{subsec:commitment} shows that 
the presence of adverse selection consumers may serve as a lock-in device, allowing for implementation of the commitment solution.
The argument relies on firms making a negative inference about consumer's risk level when observing a switching consumer.
The presence of such pessimistic beliefs is in line with the findings in \cite{cohen2005asymmetric} who shows, in the context of Israeli auto insurance, that consumers switching insurance companies have disproportionally bad accident histories and are high risk.
In a related paper, \cite{ThoronJPE2005} 
studies a two-period model with ex-ante symmetric information and learning, and shows that keeping firms from observing the accident history of switching consumers serves as a commitment device and leads to welfare improvements.


\section{\label{sec:Model}Model}

\subsubsection*{Types}
A consumer (she/her) lives for $T \leq \infty$ periods.
At the beginning of each period, she privately observes her type 
$\theta_{t}\in\Theta \equiv \left\{ l,h \right\}$,
which determines a probability distribution over realized income $y_{t}\in Y$,
with $Y \subset \mathbb{R}_+$ finite.
The occurrence of higher losses or damages is represented by a lower level of final income.\footnote{
    If the customer has fixed per period flow income $y_{0}>0$ and losses $\mathfrak{l} \in \mathfrak{L}$, her realized income is 
    $y=y_{0}-\mathfrak{l}$.
} 

I refer to type $h$ as the high-type.
In each period $t=1,\dots,T$, the probability distribution of income level $y_t$ depends on type $\theta_t$ and is represented by $p_i \in \Delta Y$, for $i=l,h$, satisfying
\[
\sum_{y\in Y}p_{{l}}\left(y\right)y<\sum_{y\in Y}p_{{h}}\left(y\right)y.
\]

I assume types $\left\{ \theta_t \right\}_{t=1}^T$ follow Markov process.
The time-invariant transition probabilities are denoted as
\[
    \pi_{ij} \equiv \mathbb{P}\left(\theta_{t+1}=j\mid\theta_{t}=i\right),
\]
while the distribution of initial type $\theta_1$ is denoted by 
$\pi_i \equiv \mathbb{P}(\theta_1 = i)$.

Types are persistent, i.e., having a given type in period $t<T$ leads to a higher probability of having the same type in period $t+1$. In short, we assume that 
$\pi_{ii} > \pi_{ji}$, for $i,j \in \left\{ l,h \right\}$.
Note that $\pi_{ll}=\pi_{hh}=1$ is included as a special case.\footnote{
    The time-invariance assumptions made here can be significantly relaxed, as discussed in Section \ref{sec:conclusion}.
}

\subsubsection*{Preferences}

Consumer preferences over final consumption flows are determined by Bernoulli utility function 
\(
u:\mathbb{R}_{+} \mapsto \mathbb{R},
\),
 assumed to be twice continuously differentiable, strictly concave
and strictly increasing.\footnote{
    This formulation rules out the relevant case of logarithmic utility, as its domain excludes zero. For finite $T$, the results extend to preferences satisfying 
    $\lim_{c\downarrow0}u\left(c\right)=-\infty$.
} 
The inverse of utility function $u\left( \cdot \right)$ is denoted as $\psi\left( \cdot \right)$.
The consumer discounts the future according to factor $\delta\in\left(0,1\right)$.
The utility obtained from deterministic consumption stream 
$\left\{ c_t \right\}_{t=1}^T$
is given by
\[
\sum_{t=1}^{T}\delta^{t-1}u\left(c_{t}\right).
\]

I study the profit maximization of a risk-neutral firm with the same discount factor. 
The payoff (or profit) obtained by a firm is determined by its net payments made to the consumer, if its contract is accepted, and zero otherwise. 
Hence the payoff obtained by a firm, for a given realization of the income and consumption paths 
$\left\{ y_t , c_t \right\}_{t=1}^T$
is given by
\[
    \sum_{t=1}^{T}\delta^{t-1}\left(y_{t}-c_{t}\right).
\]
Both the firm's and consumer's preferences are extended to random outcomes by using expected payoffs.

\subsubsection*{Flow contracts}

A flow contract is a standard single period insurance policy: it specifies
a premium and coverage for all possible income realizations. I assume
that the income realization is observable and contractable and hence
the fulfillment of a flow contract does not involve incentive considerations.
For tractability, a flow contract is described here through the induced
final consumption of the consumer, i.e., the set of flow contracts is
given by 
\(
Z\equiv\mathbb{R}_{+}^{Y}.
\)


Each contract $z\in Z$ specifies that the final consumption of the
consumer if income $y\in Y$ is realized, which is equal to the income
$y$ plus any policy coverage minus the premium paid, is equal to
$z\left(y\right)$. 

If a flow contract $z$ is provided to a consumer with type $\theta$, the consumer's utility from flow contracts can be
described by function 
\[
v\left(z,\theta\right)\equiv\sum_{y\in Y}p_{\theta}\left(y\right)u\left[z\left(y\right)\right],
\]
while the profits obtained by a firm can be described by
\[
\xi\left(z,\theta\right)\equiv\sum_{y\in Y}p_{\theta}\left(y\right)\left[y-z\left(y\right)\right].
\]

\subsubsection*{Long-term contracts}

A long-term contract is a mechanism which specifies at the initial period $t=1$ the flow contracts to be offered to the consumer at each period, which may depend on messages sent by the consumer and, potentially, information about the history of past income realizations.

In several insurance markets, regulatory restrictions limit the extent to which firms can explicitly use the history of accidents or losses in pricing contracts (see \cite{handel2015equilibria}, for example).
I allow for such restrictions by assuming that firms are only able to use a coarsening of the history of past income realizations. 
These restrictions are represented by a \textit{signal structure}, 
which is composed of a finite \textit{set of signals} $\Phi$ and a surjective \textit{signal function}
$\phi:Y \mapsto \Phi$. 
The signal structure is exogenously fixed. 
It imposes restrictions on the firm's contract design problem, as the income realization $y_t$ can only impact future offers via $\phi_t \equiv \phi(y_t)$.
 
One special case is that of \textit{fully-contingent mechanisms}, with $\Phi=Y$ and 
$\phi\left(y\right)=y$ for all $y\in Y$.
In this case, prices depend explicitly on both the history of reports by the consumer as well as the observed history of accidents, both of which are informative to firms.
Another extreme case of interest is that of \textit{realization independent mechanisms}, which corresponds to $\Phi$ being a singleton. In this case, firms are completely unable to use the history of previous income realizations in pricing.
While the results in Subsections \ref{subsec:realization-independent-contracts} and \ref{subsec:commitment} focus on the case of realization-independent mechanisms, all other results apply for an arbitrary signal structure.

From the revelation principle, we can restrict attention to direct mechanisms, where the set of possible messages in each period coincides with the set of types $\Theta$, and truthful equilibria, in which the consumer finds it optimal to truthfully report her type. 
I denote the history of announcements and signals up to period $t$ as the observable history
$\eta^{t}\in H^{t}\equiv\Phi^{t}\times\Theta^{t}$.\footnote{
    For completeness, we define $H^{0}\equiv\left\{ \emptyset\right\} $.
}

I denote a history of types $(\theta_1,\dots,\theta_t)$ up to period $t$ as $\theta^t$, the expanded history $(\theta_1,\dots,\theta_t,\theta_{t+1})$ as $(\theta^t,\theta_{t+1})$, the sub-history $(\theta_\tau,\dots,\theta_{\tau'})$, for $\tau\leq \tau' \leq t$, as 
$\left[ \theta^t \right]_\tau^{\tau'}$, 
and, for a history $\theta^t$ and $\tau \leq t$, refer to $\theta_\tau$ as
$\left[ \theta^t \right]_\tau$. 
Also define $h^t$ as the $t$-period history $(h,\dots,h)$.
The same notation is used for income realizations $y_t$, signal realizations $\phi_t$ and histories $\eta^t$.

\begin{figure}[htbp]
    \centering
    \vspace*{3em}
    \includegraphics[scale=0.3, trim=200 0 200 200]{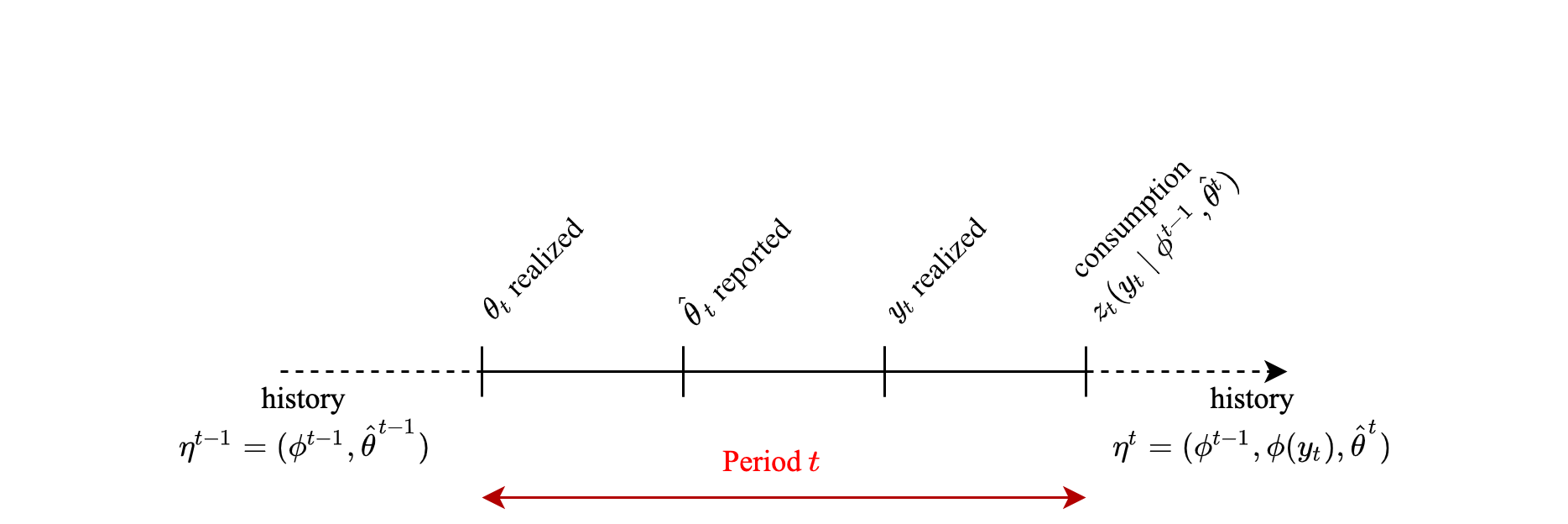}
    \vspace*{2em}
    \caption{Timing of events of a particular period $t$ within a direct mechanism}
    \label{fig:TimingMechanism}
\end{figure}

A direct mechanism is defined as  $M=\left\{ z_{t}\right\} _{t=1}^{T}$, with $z_{t}:H^{t-1}\times\Theta\mapsto Z$, and the set of direct mechanisms is $\mathcal{M}$. 
A direct mechanism 
$M=\left\{ z_{t}\right\} _{t=1}^{T}$
specifies the flow contract 
$z_{t}\left(\eta^{t},\hat{\theta}_{t}\right)$
to be provided to the consumer at each period $t$, which depends on the history of signals and announcements up to period $t-1$, as well as the new report $\hat{\theta}_{t}$ made at period $t$. The flow contract at period $t$ determines the level of coverage obtained by the consumer within period $t$, and hence her  realized consumption in period $t$, which is given by
\[
    z_{t}\left(y_{t}\mid\eta^{t},\hat{\theta}_{t}\right)
\] 
and also depends on the realized income $y_{t}$. 
Figure \ref{fig:TimingMechanism} illustrates the sequence of events within a direct mechanism for a particular period.
I assume that flow contracts with arbitrary income-contingent transfers can be executed without frictions.
In the example of health insurance, this means that
can specify and enforce transfers/coverage that depends directly on health shocks within a given period, but may --- depending on $\phi$ --- not be able to use this information explicitly when determining future offers to be made to the consumer.

A private history in period $t$ includes all information available to the consumer, and is described as 
$ \eta_{p}^{t}=\left(y^{t} , \hat{\theta}^{t} , \theta^{t}\right)\in H_{p}^{t} $,
which includes the history of income realizations $y^{t}$, type realizations $\theta^{t}$ and reported types $\hat{\theta}^{t}$.
The set of private histories in
period $t$ is denoted as $H_{p}^{t}$. A reporting strategy is denoted
by $r=\left\{ r_{t}\right\} _{t=1}^{T}$ with $r_{t}:H_{p}^{t-1}\times\Theta\mapsto\Theta$.
The truth-telling strategy is denoted by $r^{*}$ and satisfies $r_{t}^{*}\left(\eta_{p}^{t-1},\theta\right)=\theta$,
for all $\eta_{p}^{t-1}\in H_{p}^{t-1}$ and $\theta\in\Theta$. 
The history of reports up to period $t$ generated on-path by strategy $r$ --- which depends solely on realized history types $\theta^{t}$ and income levels $y^{t-1}$ ---
is denoted as 
$r^{t}\left( y^{t-1} , \theta^{t}\right)$. 
The set of reporting strategies
is denoted as $R$. The consumer's payoff from a reporting strategy $r\in R$
is denoted as 
\[
V_{0}^{r}\left(M\right)
\equiv
\mathbb{E}\left\{ 
    \sum\delta^{t-1}v\left[
        z_{t}\left( \phi^{t-1} , r^{t}\left( y^{t-1} , \theta^{t} \right)\right)
        ,\theta_{t}
    \right]
\right\} .
\]

\begin{defn} \label{def:incentive-compatility}
A direct mechanism $M=\left\{ z_{t}\right\} _{t=1}^{T}$ is incentive
compatible if, for all $r\in R$,
\[
V_{0}^{r^{*}}\left(M\right)\geq V_{0}^{r}\left(M\right).
\]
\end{defn}
The set of incentive compatible mechanisms is denoted as $\mathcal{M}_{IC}$.
Finally, for an observable history $\eta^{t-1}=\left(\theta^{t-1},y^{t-1}\right)\in H^{t-1}$
and period $t$ type $\theta_{t}\in\Theta$, 
we denote the continuation
utility of a consumer following reporting strategy $r$ as 
$V_{t}^r \left(M\mid\eta^{t-1},\theta_{t}\right)$,
or simply as
$V_{t}\left(M\mid\eta^{t-1},\theta_{t}\right)$
for reporting strategy $r^{*}$.
When the mechanism $M$ is clear --- such as the profit maximizing one --- we may simply use
$V_t (\eta^{t-1}, \theta_t)$
for brevity.

\section{Profit maximization} \label{Sec:profit-maximizing-contracts}

I assume throughout that the consumer privately learns $\theta_1$ prior to contracting with the firm.
An insurance firm's mechanism design problem can be separated into two parts. 
First, a firm must decide on how attractive its offer is for different types of potential customers, which is described by the total expected utility from participation (utility choice). 
Second, firms must choose the mechanism details in order to maximize profits within all mechanisms that deliver the same utility to the consumer (feature design). 
I start by focusing on the feature design problem and provide a characterization of profit maximizing mechanisms, for fixed discounted expected utility to be provided to the consumer with each initial type.
The study of the feature design problem allows for the characterization of qualitative properties of optimal mechanisms that hold under multiple market structures which lead to different equilibrium utility levels for the consumer. 
Sections \ref{sec:competitive-analysis} and \ref{sec:monopoly} analyze the cases of perfect competition and monopoly, respectively.

The total discounted expected profits from a direct mechanism
$M\in\mathcal{M}$ is given by 
\begin{equation}
\Pi\left(M\right)\equiv\mathbb{E}\left[\sum_{t=1}^{T}\delta^{t-1}\xi\left(z_{t}\left(\eta^{t-1},\theta_{t}\right),\theta_{t}\right)\right].\label{eq:define_profit}
\end{equation}

I am interested in studying market settings where consumers receive
offers by firms after learning their initial state $\theta_{1}\in\left\{ l,h\right\} $.
The set of feasible utility pairs for both initial types consistent with finite profits is given by\footnote{
    Restricting attention to mechanisms generating finite profits is a technical condition that avoids the analysis of transversality conditions when studying the firm's problem and is without loss for the applications considered in Sections \ref{sec:competitive-analysis} and \ref{sec:monopoly}. It is innocuous in the case $T < \infty$.
} 
\[
    \mathcal{V}\equiv\left\{
        \left(V_{1}\left(M\mid l\right),V_{1}\left(M\mid h\right)\right)
        \mid 
        M\in\mathcal{M}_{IC},\Pi(M) > - \infty
    \right\}.
\]

I refer to the profit maximization problem of the firm, for any $V=\left(V_{l},V_{h}\right)\in\mathcal{V}$
as 
\begin{equation}
\Pi^{*}\left(V\right)\equiv\sup_{M\in\mathcal{M}_{IC}}\Pi\left(M\right),\label{eq:profit_max}
\end{equation}
subject to, for $\theta\in\left\{ l,h\right\} $,
\begin{equation}
V_{1}\left(M\mid\theta\right)=V_{\theta}.\label{eq:utility_constraints}
\end{equation}

If this problem has a unique solution, we denote it as $M^V$.

\section{Complete information benchmark}

In the absence of private information, efficient long-term contracts provide complete coverage to the consumer, with final consumption that does not depend on their realized history of incomes and types, except for initial type $\theta_1$.
I denote a full coverage flow contract with consumption level
$c\in\mathbb{R}_{+}$ as $z^{c}$, i.e.,
\[
z^{c}\left(y\right)=c,
\]
for all $y\in Y$.

I define the complete information problem as that of maximizing profits,
given by (\ref{eq:define_profit}), subject to payoff constraint (\ref{eq:utility_constraints}),
with choice set $\mathcal{M}$. The solution to the complete information
problem is denoted by $\left\{ z_{t}^{CI}\right\} _{t=1}^{T}$.
The following lemma, whose proof is standard and hence omitted, formally states that the complete information solution is perfect consumption smoothing.

\begin{lem}
(Complete information benchmark) The solution to the full information
problem satisfies
\[
z_{t}^{CI}\left(\eta_{p}^{t-1},{\theta}_{t}\right)=z^{c^{CI}\left(\theta_{1}\right)},
\]
where $c^{CI}\left(\cdot\right)$ is defined by 
\[
    \sum_{t=1}^{T}\delta^{t-1}u\left(c^{CI}\left(\theta_1\right)\right)
    =
    V_{\theta_1}.
\]
 
\end{lem}

If $V_l = V_h$, the complete information solution is incentive compatible and hence solves $\Pi^*(V)$.
Our analysis of distortions will focus on utility pairs $V \in \mathcal{V}$ satisfying 
$V_h > V_l$,
i.e., where the utility delivered to an initially high-type consumer is higher than that of a low type. The focus on this case is justified in Sections \ref{sec:competitive-analysis} and \ref{sec:monopoly}, where this inequality is shown to hold in equilibrium for both the competitive and monopoly settings.
In the reverse case where $V_l > V_h$, all the results presented hold when interchanging the role of types $h$ and $l$. In particular, the relevant relaxed problem would consider ``upward'' incentive constraints.

\section{Incentives and distortions} \label{sec:Incentives_Distortions}

I now characterize the solution to the profit maximization problem (\ref{eq:profit_max}).
The main result in this section, Lemma \ref{Lem:Properties_relaxed_problem}, provides a characterization of the solution of the relaxed problem, which is used to show that the solution to the relaxed problem also solves the firm's original profit maximization problem.

In an optimal mechanism, the impact of any information revealed over time 
--- be it signal realizations or type announcements ---
is determined by its likelihood ratio: the ratio of its probability when coming from either a low- or high-type. 
For example, income realizations that are relatively more likely for high-types are rewarded with higher flow and higher continuation utilities. 
Additionally, the assumption of type persistence implies that 
future high-type announcements, under truth-telling, are relatively more likely to come from a consumer with initially high-type. As a consequence, high-type announcements are rewarded with more attractive continuation contracts.

More precisely, a solution to the relaxed problem is shown to satisfy three monotonicity properties that provide insights into the the efficient provision of dynamic incentives. These monotonicity properties are also used, in Subsection \ref{subsec:sufficiency}, to guarantee that the solution to the relaxed problem solves the original profit maximization problem.

\subsection{Relaxed problem} \label{subsec:relaxed-problem}

Define a reporting strategy $r$ to be a one-shot critical deviation
(OSCD) from truth-telling at period $t$ with signal history $\phi^{t-1}$
if the consumer reports a high-type when having a low-type for the \textit{first} time at period $t$, but otherwise reports her type truthfully.
In other words, it satisfies 
$r_t (\phi^{t-1},h^{t-1},h^{t-1},l)=h$, 
for some period $t \in \left\{ 1,\dots,T \right\}$ and signal history $\phi^{t-1} \in \Phi^{t-1}$,
but is otherwise identical to the truth-telling strategy.

I say that a mechanism is one-shot incentive compatible (OSIC) if
the consumer has no profitable OSCD, and the set of OSIC direct mechanisms
is defined as $\mathcal{M}_{OSIC}$.

\begin{figure}[htbp]
    \centering
    \vspace*{2em}
    \includegraphics[scale=0.25]{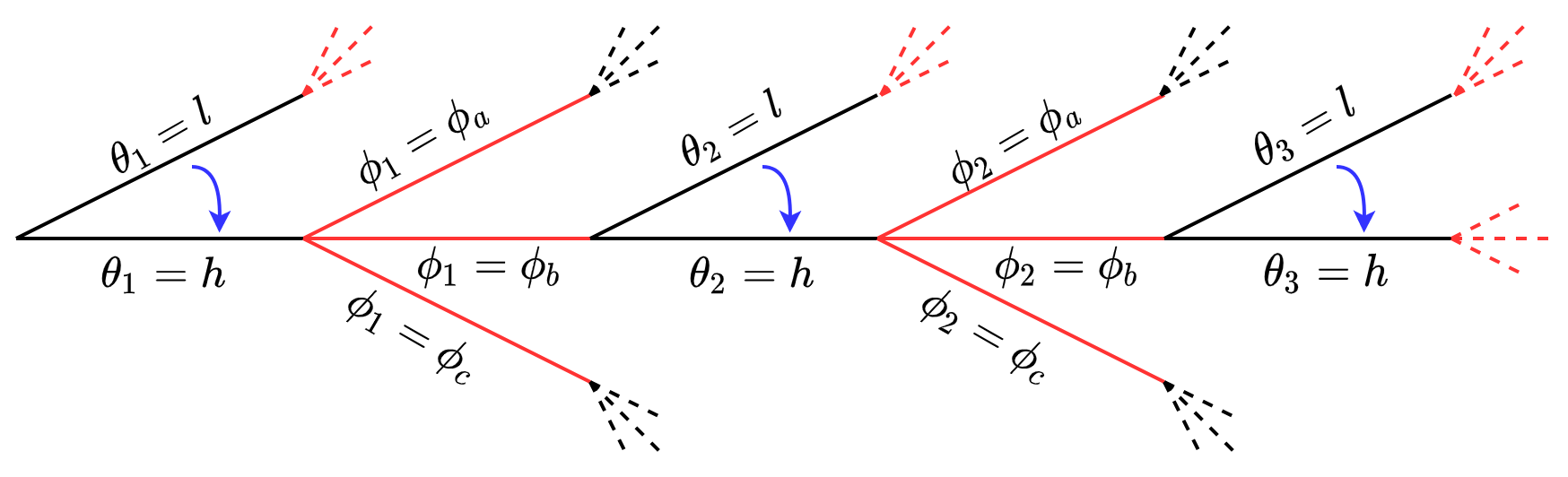} 
    \caption{This figure represents the set of binding incentive constraints in the relaxed problem. The black branches represent possible type realizations, red branches correspond to possible signal realizations, and the blue curved arrows represent the upward incentive constraints considered in the relaxed problem.} \label{fig:OneShotIncentiveConstraints}
\end{figure}

One-shot deviations are restrictive in three ways. It involves a misreport
in a single contingency, and truth-telling otherwise. This single
deviation occurs along a truth-telling path, i.e., the consumer considers
a single misreport at a given period $t$ after $t-1$ periods of
truth-telling. Finally, this deviation involves the consumer announcing
to have a low-risk type within period $t$, when in fact having a high-risk type. 
Figure \ref{fig:OneShotIncentiveConstraints} represents graphically the potential deviations considered in the relaxed problem if the set of possible signals is $\Phi = \left\{ \phi_a, \phi_b, \phi_c \right\}$. The blue arrows representing the direction of misreports, where the low-type pretends to have high-type in a given period.

The relaxed problem is the following:
\begin{equation}
\Pi^R\left(V\right)\equiv\sup_{M\in\mathcal{M}_{OSIC}}\Pi\left(M\right),
\label{eq:RelaxedProblem}
\end{equation}
subject to, for $\theta\in\left\{ l,h\right\} $,
\[
V_{1}\left(M\mid\theta\right)=V_{\theta}.
\]

\begin{assum}
    The relaxed problem has a solution, for all $V \in \mathcal{V}$.
\end{assum}

As shown below, existence of a solution is guaranteed in the presence of a finite horizon ($T<\infty$), or in the infinite horizon case as long as utility function $u(\cdot)$ is bounded. 
I do not provide an exhaustive analysis of the issue of existence of a solution for the infinite horizon case, 
but the characterization results in this paper apply more generally whenever an optimal mechanism exists.

\begin{lem}
    If $T<\infty$ or $u(\cdot)$ is bounded, the relaxed problem has a solution for all $V \in \mathcal{V}$.
\end{lem}

\begin{proof}
    Consider any $V \in \mathcal{V}$. 
    The definition of $\mathcal{V}$ implies that $\Pi^R(V) \geq \Pi^*(V) > -\infty$.
    Consider a sequence of mechanisms 
    $\left\{ M^n = \left\{ z^n_t \right\}_{t\leq T} \right\}_{n \in \mathbb{N}}$ feasible in problem $\Pi^R(V)$ such that $\Pi(M^n) > - \infty $ and
    $ \Pi(M^n) \rightarrow^{{n \rightarrow \infty}} \Pi^R(V)$.
    The fact that $ \Pi(M^n) > - \infty$ implies that, for each period $t$ and history 
    $(\phi^{t-1},\theta^{t}) \in \Phi^{t-1}\times \theta^t$, sequence
    $\left\{ z^n_t(\phi^{t-1} , \theta^t) \right\}_{n \in \mathbb{N}}$
    is uniformly bounded by a constant
    $\bar{z}(\phi^{t-1},\theta^{t}) < \sup_{c \geq 0}u(c)$
    and, as a consequence, we can assume without loss (potentially using a subsequence) that 
    $\left\{ z^n_t(\phi^{t-1} , \theta^t) \right\}_{n \in \mathbb{N}}$ 
    converges for any $t,\phi^{t-1},\theta^t$.
    Now let $M^* \equiv \left\{ 
        \lim_n z^n_t
     \right\}_{t \leq T}$.
    If $u(\cdot)$ is bounded or $T<\infty$, the dominated convergence theorem implies that $V^r(M^*) = \lim_{n \rightarrow\infty} V^r(M^n)$ for any reporting strategy $r$, which implies that $M^*$ is feasible in problem $\Pi^R(V)$.
    Fatou's lemma implies that (the profits of any two mechanisms only differ in the consumption paid out to the agent):
    \begin{equation}
        -\Pi(M^*) \leq \liminf_n \left[ -\Pi(M^n) \right] = -\Pi^R(V),
    \end{equation}
    which means that $M^*$ solves $\Pi^R(V)$.
\end{proof}

Define $\tau (\eta ^ {t-1}, \theta_t)$ as the first period at which the consumer has a low type (if any), i.e.,
\[
\tau (\eta ^ {t-1}, \theta_t) = \min \{
    t'  \in \{0, \dots , t \} \mid [\theta^ {t-1}, \theta_t]_{t'} = l
\},
\]
with $\inf \emptyset = \emptyset$. The example 
$ \tau
\left( 
    \phi^t,(l,h,\dots,h)
\right)
=1
$
represents a first period low-type, and $\tau = \emptyset$ represents the absence of a low-type realization, i.e., 
$ \tau
\left( 
    \phi^t,(h,\dots,h)
\right)
= \emptyset
$.
In the rest of this section, we omit the dependence of $\tau$ on the history of signals and types for brevity.

For any income level $y \in Y$ and signal $\phi_0 \in \Phi$, define likelihood ratios
\[
    \ell (y) \equiv \frac{p_{l}\left(y\right)}{p_{h}\left(y\right)},
\]
and
\begin{equation*}
    \ell (\phi_0) \equiv \frac{
        \sum_{y \in \phi^{-1}(\phi_0)} p_{l}\left(y\right)
    }{
        \sum_{y \in \phi^{-1}(\phi_0)} p_{h}\left(y\right)
    }.
\end{equation*}

As is standard in contract theory, these ratios are useful because they represent how useful each  income/signal realization is in screening different types. 
A realization with high likelihood ratio is more ``indicative'' of a low-type.

The following Lemma states that the relaxed problem has a unique solution and outlines key properties of its solution.
This result is proved in the Appendix and its proof exploits the recursive structure of problem $\Pi^R$.

\begin{lem} \label{Lem:Properties_relaxed_problem}
For any $V\in\mathcal{V}$, the relaxed problem has a unique solution.
Moreover, if $V_{h}>V_{l}$, its solution satisfies:

(i) All OSIC constraints hold as equalities.

(ii)  No distortions following low-type: there exists 
$\{ c_t \}_{1}^T$,
with $c_t:\Phi^{t-1} \mapsto \mathbb{R}_+$
such that 
\[
    z_t (\phi^{t-1},\theta^{t-1}, \theta_t) = z^{c_\tau (\phi^{\tau-1})},
\]
whenever $\tau \neq \emptyset.$

Additionally, there exist $\{\mu_t, \lambda_t\}_{t=1}^{T}$ with 
$(\mu_t, \lambda_t):\Phi^t \mapsto \mathbb{R} \times \mathbb{R}_+$
such that,
for any $(\eta ^ {t-1}, \theta_t)$ such that $\tau = \emptyset$,

(iii) Flow contracts following high-types have partial coverage: 
\begin{equation} \label{FOC_relaxed}
       \mu_{t-1}(\phi^{t-1}) - \lambda_{t-1}(\phi^{t-1})  \ell(y_t)
        \leq \frac{1}{ u' \left( z_t ( y_t \mid \eta ^ {t-1}, h) \right)} 
\end{equation}
with (\ref*{FOC_relaxed}) holding as a equality if $z_t(y_t \mid \eta ^ {t-1}, \theta_t) > 0$.

(iv) Future type reward: 
\[
    V_{t}(\eta^{t-1},h) > V_{t}(\eta^{t-1},l)
\]

(v) Future signal effect:
for any
$\phi, \phi'$ such that $\ell (\phi') \leq \ell (\phi)$: 
\begin{multline*}
    \pi_{lh} V_t( (\phi^{t-2},\phi' , h^{t-1}) , h) 
    +\pi_{ll} V_t( (\phi^{t-2},\phi' , h^{t-1}) , l) \\
    \geq 
    \pi_{lh} V_t( (\phi^{t-2},\phi , h^{t-1}) , h)
    +\pi_{ll} V_t( (\phi^{t-2},\phi , h^{t-1}) , l)    
\end{multline*}

\end{lem}

Property (i) state that all upward incentive constraints considered in the relaxed problem hold as equalities. 
The optimal contract is supposed to provide higher utility to a consumer with high initial type, while deterring deviations from low-types.
Given the presence of type persistence, one way to achieve this goal is to use a \textit{future} high-type realization as a signal that the consumer's initial type is $\theta_1 = h$. In other words, the utility gap between consumers with initially high and low types is propagated to future periods, with consecutive high-type announcements being ``rewarded'' with higher utility. 
For this reason, upward incentive constraints bind not only at $t=1$, but in all periods.

Property (ii) corresponds to a ``no distortion at the bottom'' result. This follows directly from the fact that the relaxed profit maximization problem ignores ``downward'' incentive constraints. 
In Subsection \ref{subsec:sufficiency} we show that those ignored constraints are guaranteed to hold in the solution to the relaxed problem.

Properties (iii)-(v) of Lemma \ref{Lem:Properties_relaxed_problem} illustrate how exogenous signal information ($\phi_t$) and endogenous reports ($\hat{\theta}_t$) are used in an optimal mechanism to efficiently screen consumers with different types using both within-period coverage as well as continuation utility.
Given the crucial role of this properties, we now discuss in detail the interpretation of each such property and show, in Lemmas \ref{Lem:Flow_Monotonicity} and \ref{Lem:Continuation_Monotonicity}, how these properties imply that incentives within the mechanism are properly aligned. These results are akin to the standard role played by the monotonicity property in mechanism design, which are used to show that mechanisms solving an appropriately defined relaxed problem ignoring certain incentive constraints satisfies the ignored incentive constraints.

\subsection{Monotonicity} \label{sec:monotonicity}

I now introduce a new set of monotonicity conditions on mechanisms which, together with binding ``upward'' incentive compatibility constraints, imply incentive compatibility. 
Lemma \ref{Lem:Properties_relaxed_problem} implies that the solution to the relaxed problem satisfies all three monotonicity conditions and all OSIC constraints as equalities, and hence solves the firm's original problem.

As discussed in the \hyperref[Sec:Appendix]{Appendix}, these monotonicity properties allows us to exploit the recursive structure of the firm's relaxed problem, which can be broken into a series of one-period problems in which the designer chooses in each period a pair of flow contracts as well as promised continuation utilities to the agent.

In this subsection, I discuss how the monotonicity conditions provide insights into the the efficient provision of dynamic incentives and their role in connecting the relaxed and original profit maximization problems.

\subsubsection{Flow monotonicity}

Flow monotonicity is akin to standard static monotonicity notions, in that it guarantees that within-period incentives are aligned by making sure that the partial coverage contract tailored to the (current) high-type consumer rewards accident realizations that are indicative of a high type, according to a likelihood ratio condition.

More precisely, flow monotonicity corresponds to property (iii) in Lemma \ref{Lem:Properties_relaxed_problem}.
For any income level $y \in Y$, the likelihood $\ell (y)$ provides a measure of how much income $y$ is indicative of low-types.
As a consequence, it is expected that profit maximizing contracts seeking to attract high-type consumers will use flow contracts that punish income realizations that are indicative of low-types (high $\ell (y)$) and reward realizations that are indicative of high-types (low $\ell (y)$).

\begin{defn}
    A flow contract $z \in Z$ satisfies flow monotonicity if it satisfies
    \begin{equation*}
        \ell (y') > \ell (y) \implies z(y') \leq z(y).
    \end{equation*}
\end{defn}

The use of flow contracts satisfying flow monotonicity implies that consumers with high-types have a higher benefit from choosing them, relative to a full-coverage contract.

\begin{lem} \label{Lem:Flow_Monotonicity}
    For any pair of flow contracts $z,z^c \in Z$, with $z$ satisfying flow monotonicity and $c\geq 0$,
    \begin{equation*}
        v(z , h) - v(z^c , h) \geq v(z , l) - v(z^c , l)
    \end{equation*}
\end{lem}

\begin{proof}
    First notice that $v(z^c , h) = v(z^c , l)$.
    Consider an ordering
    $\left\{ y_{[k]} \right\}_{k=1}^{\# Y}$
    of income realizations such that $\ell (y_{[k]})$ is weakly decreasing in $k$.
    Flow monotonicity implies that both $z(y_k)$ and 
    $\left[ 1 - \ell (y_{[k]}) \right]$ 
    are increasing in $k$. 
    Hence we have that
    \begin{align*}
        v(z , h) - v(z , l) & = \sum_{k=1}^{K} p_h (y_k) u(z(y_k)) \left[ 1- \ell (y_k) \right]\\
                            & \geq  \left\{ \sum_{k=1}^{K} p_h (y_k) u(z(y_k)) \right\}  
                            \left\{ \sum_{k=1}^{K} p_h (y_k) \left[ 1- \ell (y_k) \right] \right\},
    \end{align*}
    which is equal to zero since the last term  in this expression is zero.
\end{proof}

\subsubsection{Continuation monotonicity}

The notion of flow monotonicity is inherently static.
However, in a dynamic environment firms have extra instruments to screen consumers, namely they can use future contracts as additional screening instruments. 
From the consumer's incentive perspective, the use of such dynamic incentive schemes are represented by changes to their expected continuation utility within the mechanism as a response to either their choices or realized income within a period.
The term continuation monotonicity relates to how, in any given period, the consumer's continuation utility depends on her previous announcements
and signal realizations,
which correspond to respectively to properties (iv) and (v) in Lemma \ref{Lem:Properties_relaxed_problem}.

The first such notion is continuation-signal-monotonicity (CSM), which states that signal realizations that are indicative of low-types are punished in future periods. From the point of view of period $t$, the consumer's continuation payoff from period $t+1$ onwards depends no only on period $t$'s realized signal but also on her realized type in period $t+1$. CSM compares different signal realizations, while taking averages over possible future types using the probability distribution of a low-type in period $t$, who should be discouraged from misreporting their type in period $t$.

\begin{defn}
    For any $t<T$, a mechanism $M$ satisfies CSM at period $t$ and history $\eta^{t-1}$ if, for any $y,y' \in Y$
    \begin{equation*}
        \ell (y') > \ell (y)
    \end{equation*}
    implies
    \begin{multline*}
        \pi_{lh} V_t( M \mid (\phi^{t-2},\phi' , h^{t-1}) , h) 
        +\pi_{ll} V_t( M \mid (\phi^{t-2},\phi' , h^{t-1}) , l) \\
        \geq 
        \pi_{lh} V_t( M \mid (\phi^{t-2},\phi , h^{t-1}) , h)
        +\pi_{ll} V_t( M \mid (\phi^{t-2},\phi , h^{t-1}) , l)    
    \end{multline*}
\end{defn}

In a similar vein to the previous notions of monotonicity, we now introduce the notion of continuation-type-monotonicity (CTM), which considers the impact of the consumer's type on their continuation utility. 
As discussed in Section \ref{Sec:profit-maximizing-contracts}, we restrict attention to mechanisms that deliver a higher discounted utility to a consumer with initial type $\theta_1=h$. 
CTM requires that this ordering holds for a specific period and history.

\begin{defn}
    For any $t<T$, a mechanism $M$ satisfies continuation type monotonicity in period $t$ and history $\eta^{t-1}$ if, 
    for any $\eta^t= (\eta^{t-1}, \phi, h)$,
    \begin{equation*}
        V_{t+1}(M \mid \eta^{t},h) > V_{t+1}(M \mid \eta^{t},l).
    \end{equation*}
\end{defn}

These monotonicity properties allows us to show that incentives are aligned since the continuation contract following an announcement $\hat{\theta} = h$ is more attractive to a consumer whose true type is indeed $\theta_t=h$.

For a fixed mechanism $M$, period $t<T$ and history 
$\eta^{t-1} = (\phi^{t-1}, \theta^{t-1})$, 
define the continuation utility obtained by a consumer with type $i \in \left\{ l, h \right\}$ announcing to have a high type as 
\begin{equation} \label{def:Vhat_continuation}
    \hat{V}_{t+1} (i \mid \eta^{t-1})
    \equiv
    \sum_{\phi \in \Phi} p_{i}(\phi)
        \left[ 
            \sum_{j=l,h} \pi_{ij} V_{t+1}
            \left( 
               (\phi^{t-1}, \phi , \theta^{t-1}, h) , j 
            \right)
         \right]
\end{equation}

\begin{lem} \label{Lem:Continuation_Monotonicity}
    If mechanism $M$ satisfies CSM and CTM in period $t$, given history $\eta^{t-1}$, then 
    \begin{equation*}
        \hat{V}_{t+1} (h \mid \eta^{t-1})
        \geq
        \hat{V}_{t+1} (l \mid \eta^{t-1}).
    \end{equation*}
\end{lem}

\begin{proof}
    Rearranging summation terms gives us:

    \begin{align*}
        \hat{V}_{t+1} (h \mid \eta^{t-1})
        -
        \hat{V}_{t+1} (l \mid \eta^{t-1}) 
        = & \sum_{\phi \in \Phi} p_h (\phi)
            (\pi_{hh} - \pi_{lh}) \left[ V_{t+1} (\eta^{t-1}, \phi, h, h) - V_{t+1} (\eta^{t-1}, \phi, h, l) \right] \\
        & + 
        \sum_{\phi \in \Phi} p_h(\phi)
        \sum_{j=l,h} \pi_{lj} V_{t+1}(\eta^{t-1} , \phi, h, j) \\
        &-
        \sum_{\phi \in \Phi} p_l(\phi)
        \sum_{j=l,h} \pi_{lj} V_{t+1}(\eta^{t-1} , \phi, h, j),
    \end{align*}

    Persistence of types and CTM implies that the first term is non-negative.
    The second and term terms can be rewritten as
    \begin{equation*}
        \sum_{\phi \in \Phi} 
        p_h(\phi) (1 - \ell(\phi))
        \left[ 
            \sum_{j=l,h} \pi_{lj} V_{t+1}(\eta^{t-1} , \phi, h, j) 
        \right],
    \end{equation*}
    which is positive since it is an average of the product of two terms that are positively correlated (from CSM), with the first term having zero expectation:
    \begin{equation*}
        \sum_{\phi \in \Phi} 
        p_h(\phi) (1 - \ell(\phi)) =0.
    \end{equation*}
\end{proof}

\subsection{Sufficiency of OSIC} \label{subsec:sufficiency}

Besides providing insights into the efficient design of incentives, the three monotonicity notions introduced also allows us to guarantee that the solution of the relaxed problem is feasible in the original profit maximization problem, and hence solves it.

\begin{lem}
    If $M$ solves the relaxed problem, the consumer has no profitable one-shot deviation from truth-telling in $M$.
\end{lem}

\begin{proof}
    Consider a period $t$ with private history $\eta_p^{t-1}=(y^{t-1} , \hat{\theta}^{t-1}, \theta^{t-1})$ and current type $\theta_t$.
    Since types follow a Markov process, the consumer's preferences over continuation reporting strategies are identical for 
    (i) private history 
    $\eta_p^{t-1}=(y^{t-1} , \hat{\theta}^{t-1}, \theta^{t-1})$
    with period $t$ type $\theta_t$,
    and (ii) private history
    $\tilde{\eta_p}^{t-1}=(y^{t-1} , {\theta^{t-1}}, \theta^{t-1})$
    with period $t$ type $\theta_t$.
    In other words, we only need to look for deviations for private histories without past misreports.
    
    If ${\theta}^{t-1}$ includes any low-type realization, the result holds trivially since the optimal mechanism allocates constant consumption from period $t$ onwards. I focus now on the case ${\theta}^{t-1} = h^{t-1}$.

    If $\theta_t = l$, a one-shot deviation is a special case of the constraints considered in the OSCD concept and hence is satisfied in any mechanism that is feasible in the relaxed problem.

    If $\theta_t = h$, using Lemma \ref{Lem:Properties_relaxed_problem}-(ii) we can represent the net gain from a one-shot deviation as
    \begin{equation} \label{eq:NetGain_OSD}
        \frac{1-\delta^{T-t+1}}{1-\delta} u(c_t (\phi^{t-1}))
        -
        \left[ 
            v(z_t(\eta^{t-1}, h)) + 
            \delta 
            \hat{V}_{t+1} (h \mid \eta^{t-1})
         \right]
        ,
    \end{equation}
where $\eta^{t-1} = (\phi^{t-1},\theta^{t-1})$ is the public history connected with the private history in focus and $\hat{V}_{t+1}$ is defined as in (\ref{def:Vhat_continuation}).

From Lemma \ref{Lem:Properties_relaxed_problem}, items (iii)-(v), we know that the optimal mechanism satisfies flow and continuation monotonicity and hence, using Lemmas \ref{Lem:Flow_Monotonicity} and \ref{Lem:Continuation_Monotonicity}, the net gain in (\ref{eq:NetGain_OSD}) is weakly lower than
\begin{equation}
    \frac{1-\delta^{T-t+1}}{1-\delta} u(c^t)
    -
    \left[ 
        v(z_t(\eta^{t-1}, l)) + 
        \delta \sum_{\phi \in \Phi} p_l (\phi) 
        \sum_{j=l,h} \pi_{lj} \hat{V}_{t+1} (l \mid \eta^{t-1})    
    \right]
    ,
\end{equation}
which is the net gain from truth-telling for a consumer with type $\theta_t = l$ relative to a misreport in period $t$.
From Lemma \ref{Lem:Properties_relaxed_problem}-(i), we know this is zero.
\end{proof}

The absence of profitable one-shot deviations is enough to guarantee that a mechanism $M=\left\{ z_t \right\}_{t=1}^T$ is incentive compatible as long as the consumer's reporting problem satisfies continuity at infinity, which is guaranteed as long as flow utilities are bounded:
\begin{equation*}
    \sup \left\{ 
        |v(z_t(\eta^{t-1}) , \theta)|
         \mid 
        t=1,\dots,T, \eta^{t-1}\in H^{t-1},\theta\in \Theta
    \right\}
    < \infty,
\end{equation*}
in which case we say that the mechanism $M$ is bounded. 
This condition is automatically guaranteed if time is finite or the Bernoulli utility function $u (\cdot)$ is bounded.
It holds more generally if consumption in the solution to the relaxed problem is uniformly bounded.

\begin{cor}
    If the solution to the relaxed problem is bounded, it is optimal.
\end{cor}
\begin{proof}
    If a profitable deviation $r$ from truth-telling $r^*$ exists in mechanism $M$, 
    boundedness of $M$ implies --- through standard arguments --- that 
    a finite profitable deviation $r'$ only involving misreports up to period $t<\infty$ also exists. Since one-shot deviations from truth-telling are not profitable, then modifying $r'$ by dictating truth-telling in period $t$, regardless of the history, is a weak improvement from $r'$. Repeated application of this argument implies that $r^*$ weakly dominates $r'$, a contradiction. Hence $M$ solves the relaxed problem and is feasible in the firm's original problem, so it is optimal.
\end{proof}

\section{Auxiliary problem} \label{sec:auxiliary-problem}

The characterization of coverage and price dynamics in an insurance setting poses two technical challenges.

First,
the flow contract space, $Z = \mathbb{R}_+^Y$, is multi-dimensional and does not have a natural notion of distortions, such as as under-provision in standard screening models
(\cite{mussa1978monopoly}, \cite{Myerson81}).
The underlying inefficiency in this model is the exposure of the consumer to risk, or partial coverage, which is introduced as a way of screening high-types. I introduce a notion of distortion that is directly tied to the source of distortions, which is the need to preclude low-type consumers from misreporting their types.

Second, the introduction of risk aversion in a dynamic screening environment leads to a non-separability issue absent from linear environments. The marginal efficiency loss of from the introduction of distortions potentially depends on the underlying utility level obtained by the consumers. As a consequence, the optimal contract jointly chooses both the intertemporal allocation of utility to be provided ot the consumer as well as the spreading of distortions over time.

I tackle both of these issues by introducing a tractable auxiliary static cost minimization problem. This auxiliary cost function is then used to study the original dynamic profit maximization problem.

\subsection{Definition}

From the point of view of incentives, flow contracts fulfill two roles: to provide a certain utility level to a consumer with currently high-type, while also discouraging misreports by a low-type consumer. 

Hence we consider, for $(\nu, \Delta) \in \mathbb{R}^2$, the problem of finding an  optimal flow contract $z \in Z$ 
satisfying a promise keeping constraint:
\begin{equation} \label{eq:Chi_PromiseKeep}   
    v(z,h) =\nu,
\end{equation}
as well as a threat-keeping constraint introducing a  penalty for misreporting $l$-type consumers
(we will later focus on the case $\Delta <0$)
\begin{equation} \label{eq:Chi_ThreatKeeping}
    v(z,l) = \nu - \Delta.
\end{equation}
I denote the set of pairs $(\nu, \Delta) \in  \mathbb{R}^2$ such that a flow contract satisfying both (\ref*{eq:Chi_PromiseKeep}) and (\ref*{eq:Chi_ThreatKeeping}) exists as $A \subset \mathbb{R}^2$.
The utility wedge $\Delta$ between consumers with different types is the source of distortions in the contract since exposing the consumer to risk is the only way to discourage low-type consumers from misreporting. I refer to it directly as the level of distortion.

Define the following auxiliary static cost minimization problem $\mathcal{P}^A$, for any $(\nu, \Delta) \in A$:

\begin{equation*}
    \chi\left(\nu,\Delta\right)
    \equiv
    \inf_{z \in Z}\sum_{y \in Y}p_{h}\left(y\right)z\left(y\right),
\end{equation*}
subject to constraints (\ref*{eq:Chi_PromiseKeep}) and (\ref*{eq:Chi_ThreatKeeping}).

\begin{lem} \label{Lem:Chi_uniqueness}
    For any $(\nu, \Delta) \in A$, problem $\mathcal{P}^A$ has a unique solution and $\chi$ is strictly increasing in both coordinates if $\Delta \geq 0$. Moreover, if the solution at $(\nu, \Delta)$ is interior then $\chi$ is twice differentiable in an open neighborhood of $(\nu, \Delta)$.
\end{lem}

\begin{proof}
    Follows directly from Lemma \ref{Lem:Appendix_Chi_Properties} in the Appendix.
\end{proof}

For any $(\nu, \Delta) \in A$, we denote the solution of $\mathcal{P}^A$ as $\zeta(\nu, \Delta)$.

In the remainder of the analysis, I assume  that this cost function is supermodular, i.e., that the marginal cost of distortions is increasing in the utility level delivered to the agent.

\begin{assum}
    (Cost supermodularity) \label{Assumption:Cost_supermodularity}
    Function $\chi(\cdot)$ satisfies
    \begin{equation*}
        \frac{\partial^2 \chi(\nu, \Delta)}{
            \partial \nu
            \partial \Delta
        }
        \geq 0,
    \end{equation*}
    whenever $\mathcal{P}^A$ has an interior solution.
\end{assum}

The supermodularity of function $\chi (\cdot)$ can be written in terms of utility function $u(\cdot )$, and is equivalent to the requirement that the coefficient of absolute risk aversion $r_u$ does not decrease with consumption ``too quickly''.
This is guaranteed if $u\left(\cdot\right)$ has non-decreasing absolute risk aversion (IARA), which includes constant absolute risk aversion (CARA) utility as a special case. It also holds for the case of constant relative risk aversion (CRRA) utility with coefficient above $1/2$.

More formally, defining the coefficient of absolute risk aversion as 
\begin{equation*}
    r_u(c) \equiv - \frac{u'' (c)}{u' (c)},
\end{equation*}
we can state the following result, which is proved in the Appendix.

\begin{lem} \label{prop:Sufficient_supermodularity}
    Assume $r_u:\mathbb{R}_{+}\rightarrow\mathbb{R}_{+}$ is differentiable,
    then the following are equivalent: \\
    (i) Assumption \ref{Assumption:Cost_supermodularity} holds, \\
    (ii)
    \(
    r_u'\left(x\right)u'\left(x\right)+2\left[r_u\left(x\right)\right]^{2}\geq0,
    \) \\
    (iii) $\psi''' (x) > 0$.
\end{lem}
\begin{proof}
    In the 
    \hyperref[proof:conditions-supermodularity]{Appendix}.
\end{proof}

\subsection{Interpretation} \label{subsec:interpretation}

The auxiliary problem provides a way to study feature of the flow contracts offered along the optimal mechanism. 
In this subsection, we discuss how properties of the cost function $\chi(\cdot)$ translate into risk exposure and consumption within the cost-minimizing flow contract.
The optimal intertemporal allocation of consumption and distortions, to be studied in Section \ref{sec:Utility-Distortion-dynamics}, is characterized in terms of the marginal costs of flow utility and distortions.

For any $(\nu, \Delta ) \in \mathbb{R} \times \mathbb{R}_+$ for which problem $\mathcal{P}^A$ has an interior solution $\zeta(\cdot)$, 
we show in the Appendix (Lemma \ref{lem:chi-derivatives-foc}) that
marginal costs relate to consumption in cost-minimizing contract according to:

\begin{equation} \label{eq:Chi_FOC_Derivatives}
    \frac{1}{
        u' (\zeta(y))
    } 
     = \chi_\nu (\nu, \Delta) 
    + \chi_\Delta (\nu, \Delta) \left[ 
        1 - \ell (y)
     \right] ,
\end{equation}
where we use notation $\chi_j(\cdot) \equiv \frac{\partial \chi(\cdot)}{\partial j}$.
The inverse marginal utility, which is strictly increasing in consumption, represents the marginal cost for the firm to deliver an additional infinitesimal amount of utility to the consumers for each specific income realization.
Given that $\chi(\cdot)$ is a convex function, an increase in flow utility $\nu$ (or in distortion $\Delta$) leads to an increase in $\chi_\nu$ (or in $\chi_\Delta$).
From (\ref*{eq:Chi_FOC_Derivatives}), we can separately
relate both marginal costs with the consumer's consumption and risk exposure.

First, the marginal cost of flow utility is related to the consumer's marginal utility through the following equation:
\begin{equation*}
    \sum_{y \in Y} p_h(y)
    \frac{1}{
        u' (\zeta(y))
    }
    = \chi_\nu (\nu, \Delta),
\end{equation*}
This is a standard inverse marginal utility optimality condition in contract theory (see \cite{rogerson1985repeated}, for example).
As I illustrate in the examples below, since the function 
$\left[ u'(c) \right]^{-1}$
is strictly increasing in consumption, an increase in $\chi_\nu$ can be seen as an overall increase in consumption, with the exact connection being dependent on the shape of utility $u(\cdot)$.

Second, the following condition relates the distortion level $\Delta$ to the responsiveness of consumption to income realizations within a period: for any $y,y' \in Y$
\begin{equation*}
    \frac{1}{
        u' (\zeta(y'))
    } 
    -
    \frac{1}{
        u' (\zeta(y))
    } 
     =
    -\chi_\Delta (\nu, \Delta) \left[ 
        \ell (y') - \ell (y)
     \right] .
\end{equation*}

Intuitively, cost minimizing flow contracts reward income realizations with low likelihood ratio $\ell(\cdot)$.
These two marginal cost equations imply that 
\begin{equation*}
    \chi_\Delta^2(\nu, \Delta)
    \sum_{\phi \in \Phi} p_h(y) \left[ 
        1 - \ell(y)
    \right]^2
    =
    \sum_{\phi \in \Phi} p_h(y) \left\{ 
        [u'(\zeta (y))]^{-1}
        -
        \sum_{y' \in Y}
        p_h(y') [u'(\zeta (y'))]^{-1}
     \right\}^2.
\end{equation*}

The right-hand side corresponds to the variance of the inverse marginal utility of the consumer.
Hence an increase in distortion $\Delta$, and hence in $\chi_\Delta$, generates a larger dispersion in the inverse marginal utility of consumer across income realizations. A larger distortion requires that the flow contract expose the consumer to more risk, so that low-risk consumer finds this contract less attractive. This condition shows that this additional dispersion is monotonically related to $\chi_\Delta (\cdot)$.
A clearer connection can be made for parametric families of utility functions.






\subsubsection*{Example: CRRA preferences}

Consider the case of CARA preferences, with 
\begin{equation*}
    u(c) = \frac{c^{1- \rho} -1}{1 -\rho},
\end{equation*}
with coefficient of relative risk aversion $\rho > 0$.

In this case, an increases in the flow utility $\nu$ is directly related to an increase in the $\rho$-th moment of the consumption distribution:
\begin{equation*}
    \chi_\nu (\nu, \Delta)
    = 
    \sum_{y \in Y} p_h(y)
    \zeta(y)^\rho,
\end{equation*}
while an increase in distortion $\Delta$ leads to a larger variance of the $\rho$-th power of final consumption:
\begin{equation*}
    \chi_\Delta^2(\nu , \Delta)
    =
    \frac{
        \sum_{y\in Y} \left[ \zeta(y)^\rho - \mathbb{E}_h(\zeta^\rho)\right]^2
    }{
        \sum_{y\in Y} p_h(y) \left[ 1 - \ell (y) \right]^2
    },
\end{equation*}
where
$\mathbb{E}_h(\zeta^\rho) \equiv \sum_{y \in Y} p_h(y)\zeta(y)$.

Two particular cases are of special interest. In the case $\rho =1$, where the utility becomes $u(c) = \log(c)$, the marginal costs $\chi_\nu$ and $\chi_\Delta$ are multiples of the mean and variance, respectively, of the within-period consumption of the consumer.

Alternatively, if $\rho=1/2$, the marginal costs $\chi_\nu$ and $\chi_\Delta$ are multiples of the mean and variance, respectively, of the within-period ex-post utility of the consumer --- $u(\zeta(y))$  --- in the cost minimizing contract $\zeta(\cdot )$. 
Hence distortion parameter $\Delta$ represents the consumer's exposure to risk, but evaluated from the point of view of her utility level.

\subsubsection*{Example: binary outcomes}

In Section \ref{sec:Utility-Distortion-dynamics}, I provide results on the dynamics of utility flow and distortions.
In the special case of binary outcomes, with $Y = \left\{ \underline{y} , \bar{y} \right\}$ satisfying $\ell (\underline{y}) > \ell (\bar{y})$, 
conditions (\ref{eq:Chi_PromiseKeep}) and (\ref{eq:Chi_ThreatKeeping}) imply that both are connected to consumer's income-contingent utility in a simple way:
\begin{align*}
    u(\zeta(\bar{y})) = \nu 
    + \frac{p_h(\underline{y})}{p_h (\bar{y}) - p_l (\bar{y})} \Delta, \\
    u(\zeta(\underline{y})) = \nu 
    - \frac{p_h(\bar{y})}{p_h (\bar{y}) - p_l (\bar{y})} \Delta.
\end{align*}

In which case $\nu$ and $\Delta$ correspond directly to the utility level and dispersion of the consumer within a period. 

\section{Distortion and consumption dynamics} \label{sec:Utility-Distortion-dynamics}

In this section, we use the auxiliary cost problem introduced in Section \ref{sec:auxiliary-problem} to shed light on the dynamics of distortions and consumption in the optimal mechanism. The firm's profit maximization is directly linked with problem $\mathcal{P}^A$ since each within-period flow contract provided by the firm must be optimal within the set of flow contracts that deliver the same utility level for truth-telling consumers (which corresponds to $\nu$) and discourages misreporting consumers by the same amount (which corresponds to $\Delta$). 

In other words, each contract solves the auxiliary problem $\mathcal{P}^A$ for a particular pair $(\nu, \Delta)$.
We can then separate the problem of the firm into two ``sub-problems'': 
one in which flow utility and distortions are allocated across periods, which we study in this section, 
and another one in which the chosen flow utility and distortion levels must be delivered in a cost-minimizing fashion, as discussed in Section \ref{sec:auxiliary-problem}. 
Our analysis now restricts attention to the case of finitely many periods.

The following proposition formalizes this separation, by showing that all flow contracts delivered in the optimal mechanism are indeed solutions to the auxiliary cost minimization problem, with suitably chosen pairs of utility flow and distortion.

Given optimal mechanism $M = \left\{ z_t \right\}_{t=1}^T$, define the following:
\begin{align*}
    \nu_t(\phi^{t-1}) & \equiv v \left( z_t(h^{t-1},\phi^{t-1},h),h \right),\\
    \Delta_t(\phi^{t-1}) & \equiv \nu_t(\phi^{t-1}) - v \left( z_t(h^{t-1},\phi^{t-1},h), l \right),
\end{align*}
which correspond respectively to the flow utility and distortion in the partial coverage contract offered in period $t$ following type announcements $h^t$.

\begin{proposition} \label{prop:Optimal_solves_Chi}
    The optimal mechanism $M=\left\{ z_t \right\}_{t=1}^T$ satisfies, for any $t=1,\dots,T$,
    \begin{equation*}
        z_t(h^{t-1},\phi^{t-1},h) = \zeta(\nu_t(\phi^{t-1}),\Delta_t(\phi^{t-1})).
    \end{equation*}
\end{proposition}

\begin{proof}
    Fix any period $t<T$ and signal history $ \phi^t $.
    For any mechanism $\tilde{M}=\left\{ \tilde{z}_t \right\}_{t=1}^T$, 
    the flow contract $\tilde{z}_t(h^{t-1},\phi^{t-1},h)$ only affects the OSIC constraints in the relaxed problem via
    \begin{equation*}
        v \left( \tilde{z}_t(h^{t-1},\phi^{t-1},h), h \right) \text{ and } v \left( \tilde{z}_t(h^{t-1},\phi^{t-1},h), l \right),
    \end{equation*}
    while it only affects the firm's profits via 
    \begin{equation*}
        \sum_{y \in Y}p_{h}\left(y\right)
        \tilde{z}_t(h^{t-1},\phi^{t-1},h)\left(y\right).
    \end{equation*}
    Hence modifying mechanism $M$ by substituting flow contract $z_t(h^{t-1},\phi^{t-1},h)$ by $\zeta(\nu_t(\phi^t), \Delta_t(\phi^t))$ still satisfies OSIC and (given uniqueness of solution in $\mathcal{P}^A$) strictly increases profits in the relaxed problem if $z_t(h^{t-1},\phi^{t-1},h) \neq \zeta(\nu_t(\phi^t), \Delta_t(\phi^t))$.

\end{proof}

From now on, we refer to
$\left\{ \nu_t,\Delta_t \right\}_{t=1}^T$ 
simply as the flow utility and distortion in the optimal mechanism following a sequence of high-type announcements.
I use a similar notation to refer to the flow per-period utility derived by the consumer following a first low-type announcement in period $t$:
\begin{equation*}
    \nu^l_t(\phi^{t-1}) \equiv
    u(c_t(\phi^{t-1})).
\end{equation*}

We can now use Proposition \ref{prop:Optimal_solves_Chi} and focus solely on the problem of intertemporal allocation of flow utility $\nu$ and distortion $\Delta$ within the optimal contract. 
The following proposition displays the optimality conditions connected with intertemporal allocation of utility and distortions.
Since all terms in Proposition \ref{prop:intertemporal_condition_general} depend on 
$\phi^{t-1}$, we omit this dependence for brevity.

\begin{proposition} \label{prop:intertemporal_condition_general}
    For any $t=1,\dots,T$ and $\phi^{t-1} \in \Phi^{t-1}$, if the solution to the cost minimization problem $\mathcal{P}^A$ has an interior solution, i.e.,
    \begin{align*}
        (\nu_t, \Delta_t)(\phi^{t-1}) & \in int(A), \\
        (\nu_t, \Delta_t)(\phi^{t-1},\phi') & \in int(A) \text{, for all }\phi' \in \Phi,    
    \end{align*}
    then the following optimality conditions hold
    \begin{equation} \label{eq:Intertemporal_util_Chi_general}
        \chi_\nu (\nu_t,\Delta_t) 
        = 
        \sum_{\phi \in \Phi}p_h(\phi) \left\{ 
            \pi_{hh}\chi_\nu[(\nu_{t+1},\Delta_{t+1})(\phi_t)]
            + \pi_{hl}\chi_\nu[(\nu_{t+1}^l (\phi_t) ,0)]
        \right\},
    \end{equation}
    \begin{equation} \label{eq:Intertemporal_delta_Chi_general}
        \begin{split}
            \chi_\Delta (\nu_t,\Delta_t) 
            = 
            \sum_{\phi \in \Phi} p_h(\phi) 
            \frac{\pi_{hh}}{\pi_{hh} - \pi_{lh}}
             \left\{ 
                 \chi_\Delta[(\nu_{t+1},\Delta_{t+1})(\phi_t)]
                 \vphantom{\frac{1}{2}}
            \right. \\
            \left.
                 +
                \pi_{hl}
                \left[ 
                    \chi_\nu[(\nu_{t+1},\Delta_{t+1}) (\phi_t)]
                    -
                    \chi_\nu[(\nu_{t+1}^l(\phi_t),0)]
                 \right]
                \vphantom{\frac{1}{2}}
            \right\}.
        \end{split}
    \end{equation}
\end{proposition}

The Proposition above relies on the use of local optimality conditions of the firm's profit maximization problem and hence assumes its solution is interior.
The interiority condition in Proposition \ref{prop:intertemporal_condition_general} holds if the optimal mechanism has strictly positive consumption flows.
Conditions (\ref{eq:Intertemporal_util_Chi_general}) and (\ref{eq:Intertemporal_delta_Chi_general}) characterize the dynamics of flow utility and distortions along the non-trivial histories that involve distortions.

Equation (\ref{eq:Intertemporal_util_Chi_general}) represents the efficient intertemporal allocation of flow utilities, or consumption. 
It states that the marginal cost of providing a high-type with higher flow utility in period $t$ must be equalized to the expected marginal cost of higher flow utility in period $t+1$. The marginal cost of flow utility coincides with the expectation of the inverse of the consumer's marginal utility (see Subsection \ref{subsec:interpretation}), which shows that this condition is a special case of the inverse Euler equation studied in dynamic allocation problems (see \cite{rogerson1985repeated}, \cite{farhi2012capital}).

Equation (\ref{eq:Intertemporal_delta_Chi_general}), on the other hand, represents the efficient intertemporal allocation of distortions.
Let's consider the problem, in period $t<T$, of discouraging a consumer with low period $t$ type from pretending to have a high type.
Due to the presence of type persistence, this can be done in two ways:
by exposing the consumer to risk in period $t$, which is represented by a larger $\Delta_t$; 
or by exposing the consumer to more risk in period $t+1$ as long as the consumer still claims to be of high type ($\theta_{t+1}=h$). 
Of course, given the convexity of the cost function $\chi(\cdot)$, the optimal mechanism uses both screening methods in a balanced way.

However, (\ref{eq:Intertemporal_delta_Chi_general}) illustrates the differences in using current versus future period distortions. 
First, notice that future distortions are only useful due to the persistence of types. As type persistence is reduced, or $(\pi_{hh} - \pi_{lh})$ is smaller, the optimal mechanism relies mostly on current period distortions for screening purposes. In the limit where $\pi_{hh} = \pi_{lh}$, the use of future distortions in screening is useless and, as a consequence, the optimal contract only features distortions in the first period.

Second, reducing the distortion in period $t$ while increasing it in period $t+1$ not only has a direct cost impact as it requires exposing the consumer to risk (this is captured by the $\chi_\Delta$ term), 
but it also implies that next period's type realization now has a larger impact on the consumer's final utility. 
This additional utility dispersion has a marginal cost which depends on the gap of the marginal cost of flow utility in period $t+1$ for both possible types (represented by the second and third terms on the right-hand side of (\ref{eq:Intertemporal_delta_Chi_general})). 

\subsection{Realization-independent contracts} \label{subsec:realization-independent-contracts}

I start by focusing on the case of realization-independent contracts, i.e., long-term contracts that do not use the history of past income realizations when determining the menu to be offered to the consumer within a period. As discussed in the introduction, this represents an extreme case of price restrictions that limit how much information can be used explicitly by firms when pricing consumption. 
This subsection also restricts attention to the finite-horizon case, i.e.,
$T < \infty$.

Since flow contracts now only depend on the history of announcements, we drop the dependence on history $\phi^t$.
In this case, equations (\ref{eq:Intertemporal_util_Chi_general}) and (\ref{eq:Intertemporal_delta_Chi_general}) become 
\begin{equation} \label{eq:Intertemporal_util_Chi_RI}
    \chi_\nu (\nu_t,\Delta_t) 
    = 
        \pi_{hh}\chi_\nu(\nu_{t+1},\Delta_{t+1})
        + \pi_{hl}\chi_\nu(\nu_{t+1}^l  ,0),
\end{equation}
and
\begin{align} \label{eq:Intertemporal_delta_Chi_RI}
    \chi_\Delta (\nu_t,\Delta_t) 
    = 
    \frac{\pi_{hh}}{\pi_{hh} - \pi_{lh}}
    & \left\{ 
         \chi_\Delta(\nu_{t+1},\Delta_{t+1})
         +
        \pi_{hl}
        \left[ 
            \chi_\nu(\nu_{t+1},\Delta_{t+1}) 
            -
            \chi_\nu(\nu_{t+1}^l,0)
         \right]
    \right\}.
\end{align}

The supermodularity guaranteed by Assumption \ref{Assumption:Cost_supermodularity} allows us to characterize the dynamic behavior of utility and distortions.
Let's focus on the intertemporal allocation between a given periods $t$ and $t+1$, following a sequence of high-type realizations $h^t$.
Given that the optimal contract rewards high-type announcements, the flow utility following an additional high-type realization in period $t+1$, $\nu_{t+1}$, is higher relative to that of a consumer with a low-type realization, $\nu^l_{t+1}$.

Together with the presence of distortions following a high-type announcement ($\Delta_{t+1}>0$) and supermodularity of $\chi(\cdot)$ (Assumption \ref{Assumption:Cost_supermodularity}), we can then conclude that

\begin{equation*}
    \chi_\nu(\nu_{t+1},\Delta_{t+1}) 
    >
    \chi_\nu(\nu_{t+1}^l,0).
\end{equation*}

From equation (\ref{eq:Intertemporal_util_Chi_RI}), we can already conclude that consecutive high-type announcements lead to an increase in the marginal cost of flow utility:
\begin{equation} 
        \chi_\nu(\nu_{t+1},\Delta_{t+1})
        >
        \chi_\nu (\nu_t,\Delta_t) 
        >
        \chi_\nu(\nu_{t+1}^l  ,0).
\end{equation}
Now, using intertemporal distortion allocation condition (\ref{eq:Intertemporal_delta_Chi_RI}), we have that
\begin{equation*}
    \chi_\Delta (\nu_t,\Delta_t) 
    > 
    \frac{\pi_{hh}}{\pi_{hh} - \pi_{lh}}
    \chi_\Delta(\nu_{t+1},\Delta_{t+1})
    >
    \chi_\Delta(\nu_{t+1},\Delta_{t+1})
\end{equation*}

In other words, profit maximization mandates that, along high-type path $h^T$, the marginal cost of flow utility is increasing while the marginal cost of distortions is decreasing.
Using convexity of $\chi$ and, once again, Assumption \ref{Assumption:Cost_supermodularity} (see Lemma \ref{lemma:single-crossing_chi}) these properties can be translated into statements about the dynamic behavior of flow utility and distortions, which are summarized in the following Proposition.

\begin{proposition} \label{prop:RI_monotonicity}
    If $V_h > V_l$, the optimal mechanism satisfies: \\
    (i) High-type utility flows increase, i.e., $\left\{ \nu_t \right\}_{t=1}^T$ is strictly increasing, \\
    (ii) Distortions decrease, i.e.,  $\left\{ \Delta_t \right\}_{t=1}^T$ is strictly decreasing.
\end{proposition}
\begin{proof}
    In the \hyperref[Proof_prop_4]{Appendix}.
\end{proof}

\subsection{Realization-dependent contracts} \label{subsec:realization-dependent}

Considering a general signal structure, we are able to use Proposition \ref{prop:intertemporal_condition_general} to extend the monotonicity result of Proposition \ref{prop:RI_monotonicity} in two cases. 
First, we use the case of constant absolute risk aversion with coefficient $1/2$ as an illustrative example. This knife-edge example simplifies the analysis since it is the only one in which the marginal costs of flow utility and distortions are separable.
We also look at the case of general preferences with two periods.

\subsubsection{Quadratic cost} \label{subsubsec:quadraticCase}

I now assume that the consumer's utility function takes the following form:
\begin{equation*}
    u(c) = 2\sqrt{c}.
\end{equation*}

This example is particularly tractable since the auxiliary cost function $\chi \left( \cdot \right)$ is 
(i) separable in $\nu$ and $\Delta$, i.e., 
\begin{equation*}
    \frac{
        \partial^2 \chi(\nu,\Delta)
    }{
        \partial \nu \partial \Delta
    }
    =0,
\end{equation*}
whenever twice differentiable, and 
(ii) quadratic, which implies that 
$\chi_\nu \left( \cdot  \right)$ and
$\chi_\Delta \left( \cdot  \right)$
are linear in $\nu$ and $\Delta$, respectively.

In this case, intertemporal optimality conditions (\ref{eq:Intertemporal_util_Chi_general}) and (\ref{eq:Intertemporal_delta_Chi_general}) in Proposition (\ref{prop:intertemporal_condition_general}) can be rewritten as statements in terms of utility flow and distortions:

\begin{equation} \label{eq:Quadratic_martingale_flow}
    \nu_t (\phi^{t-1})
    = 
    \sum_{\phi \in \Phi}p_h(\phi_t) 
    \left[ 
        \pi_{hh} \nu_{t+1} (\phi^t)
        + \pi_{hl} \nu_{t+1}^l (\phi^t)
    \right],
\end{equation}
\begin{equation} \label{eq:Quadratic_distortion}
    \Delta_t(\phi^{t-1}) 
    = 
    \sum_{\phi \in \Phi} p_h(\phi_t) 
    \frac{\pi_{hh}}{\pi_{hh} - \pi_{lh}}
        \Delta_{t+1}(\phi^t)
    +
    N
    \left[
        \nu_{t+1}(\phi^t)
        -
        \nu^l_{t+1}(\phi^t)
    \right],
\end{equation}
for constant $N \equiv \frac{\partial^2 \chi}{\partial \nu^2}>0$.

The first equation implies that flow utilities form a martingale in the optimal mechanism. In the optimal mechanism, the continuation utility following a high-type realization is always larger than the one following a low-type announcement, and the martingale property of flow utilities implies that \textit{flow utilities} in any given period $t$ are also larger following a high-type realization $\theta_t=h$, everything else equal.
Together with (\ref{eq:Quadratic_martingale_flow}) and (\ref{eq:Quadratic_distortion}), this implies that 
distortions follow a supermartingale, conditional on the type path $\theta^T$.
Additionally, the change in flow utility across periods is similar to the realization-independent case: a high-type realization leads to higher utility flow while a low-type realization leads to lower utility flows – when integrating over the possible signal realizations.
These statements are formalized below.

\begin{proposition} If $V_h > V_l$ and the optimal mechanism is interior, then: \\
    (i) distortions follow a supermartingale, conditional on $\theta^T$, i.e.,
    \begin{equation*}
        \Delta_t (\phi^{t-1})
        > 
        \sum_{\phi \in \Phi}p_h(\phi_t)
            \Delta_{t+1} (\phi^t)
    \end{equation*}
    and (ii) within period flow utilities are increasing in past periods' type announcements, i.e.,
    \begin{equation*}
        \sum_{\phi \in \Phi}p_h(\phi_t)
        \nu_{t+1} (\phi^t)
        >
        \nu_t (\phi^{t-1})
        > 
        \sum_{\phi \in \Phi}p_h(\phi_t)
            \nu_{t+1}^l (\phi^t)
    \end{equation*}
    
\end{proposition}

\begin{proof}
    First notice that the flow utility in period $t$ if the consumer's first low-type announcement is in period $t$ is strictly lower than that if the consumer has one more high-type realization in period $t$ since:
    \begin{align*}
        V_{t}(\phi^{t-1},h^{t-1},l) 
        = & 
        \nu_t^l (\phi^{t-1}) \sum_{\tau=t}^T\delta^{\tau-t}
        \\
        = &
        \nu_t(\phi^{t-1})
        +
        \delta 
        \sum_{\phi \in \Phi} p_l (\phi_t)
        \left[ 
            \pi_{lh} V_{t+1}(\phi^{t},h^{t},h)
            +
            \pi_{ll} V_{t+1}(\phi^{t},h^{t},l)
         \right] 
         - \Delta_t (\phi^{t-1}) \\
         < &
        \nu_t(\phi^{t-1})
        +
        \delta 
        \sum_{\phi \in \Phi} p_h (\phi_t)
        \left[ 
            \pi_{hh} V_{t+1}(\phi^{t},h^{t},h)
            +
            \pi_{hl} V_{t+1}(\phi^{t},h^{t},l)
         \right] 
         - \Delta_t (\phi^{t-1}) \\
         = &
         \nu_t(\phi^{t-1}) \sum_{\tau=t}^T\delta^{\tau-t}
          - \Delta_t (\phi^{t-1}).
    \end{align*}
    The first equality follows from $\nu^l_t$ being defined as the constant flow utility obtained by the consumer following a first low-type realization in period $t$. 
    The second equality corresponds to the binding incentive constraint in period $t$. 
    The inequality follows from Lemma \ref{Lem:Properties_relaxed_problem}-(iv). 
    Finally, the last equality follows from the martingale property (\ref{eq:Quadratic_martingale_flow}). Hence we conclude that $\nu_t(\phi^{t-1}) > \nu^l_t (\phi^{t-1})$, for all $t$ and $\phi^{t-1} \in \Phi^{t-1}$. 

    The result then follows directly from equations (\ref{eq:Quadratic_martingale_flow}) and (\ref{eq:Quadratic_distortion}).
\end{proof}

\subsubsection{General preferences}

Now consider an arbitrary signal structure and focus on the case of two periods.
As mentioned before, in maximizing profits the firm has an incentive to reward subsequent high-type announcements (as long as $V_h >V_l$), and to rely less on later periods' distortions in order to provide incentives.

As a consequence, the marginal costs of flow utility must be increasing,
while the marginal cost of distortions must be decreasing
following consecutive high-type announcements (which correspond to partial coverage), as shown below.
As discussed in Subsection \ref{subsec:interpretation}, the marginal cost terms $\chi_\nu$ and $\chi_\Delta$ are related to the consumption level and risk exposure in a given flow contract.
Hence the interpretation of Proposition \ref{prop:General_T2} is that the consecutive choices of partial coverage are rewarded, when averaging over signal realizations, with more consumption and less distortions.

\begin{proposition} \label{prop:General_T2}
    Assume $T=2$, the optimal mechanism satisfies
    \begin{equation*}
        \chi_\Delta (\nu_1,\Delta_1) 
        >
        \sum_{\phi \in \Phi}p_h(\phi)
            \chi_\Delta[(\nu_{2},\Delta_{2})(\phi)]
    \end{equation*}
    and
    \begin{equation*}
        \sum_{\phi \in \Phi}p_h(\phi)
            \chi_\nu[(\nu_{2},\Delta_{2})(\phi)]
        >
        \chi_\nu (\nu_1,\Delta_1) 
        > 
        \sum_{\phi \in \Phi}p_h(\phi) 
        \chi_\nu[(\nu_{2}^l (\phi) ,0) ],
    \end{equation*}
\end{proposition}

\begin{proof}
    Lemma \ref{Lem:Properties_relaxed_problem} implies that 
    \begin{equation*}
        V_2(\phi,h) = \nu_2(\phi,h) > 
        V_2(\phi,l)
        = 
        \nu^l_2(\phi),
    \end{equation*}
    which implies that
    $
    \chi_\nu[(\nu_{2},\Delta_{2})(\phi)]
    >
    \chi_\nu[\nu^l_{2}(\phi)]
    $, 
    for all signals $\phi \in \Phi$. 
    
    The result then follows directly from (\ref{eq:Intertemporal_util_Chi_general}) and (\ref{eq:Intertemporal_delta_Chi_general}).
\end{proof}

In the case of realization independent contracts, monotonicity properties of the marginal cost functions (which are guaranteed by supermodularity) allow us to translate statements about marginal costs directly into statements about the \textit{levels} of flow utility and distortion, as in Proposition \ref{prop:RI_monotonicity}.
Once we allow for richer signal structures, statements about the expectations of marginal costs cannot be translated into statements about the expectations of flow utility and distortion, except for the special case in which the marginal cost functions are linear in the flow utility and distortion pair $(\nu,\Delta)$ covered in Subsection \ref{subsubsec:quadraticCase}.

\section{Competitive analysis} \label{sec:competitive-analysis}

I now consider a competitive model that extends the analysis of \cite{rothschild1976equilibrium} and \cite{cooper1987multi} in allowing for both persistent non-constant risk types and offers containing dynamic mechanisms.
For now, I assume a single contracting stage between consumers and firms in the first period, which means that both firms and consumers can commit. 
The role of the commitment assumption and possible ways of relaxing it are discussed in Subsection \ref{subsec:commitment}.

Consider the following extensive form. 
A finite number of firms simultaneously offer a mechanism to a consumer. 
The consumer observes her initial type $\theta_{1}$ and decides which firm's mechanism to accept, if any. I assume exclusivity, i.e., the consumer can choose at most one mechanism.
If the buyer does not accept any offer, she obtains no insurance coverage and gets discounted utility
\begin{equation} \label{eq:OutsideOption}
    \underline{V}_i 
    \equiv 
    \mathbb{E} \left[ 
        \sum_{t=1}^T \delta^{t-1} u \left( 
            {y}_t
        \right)  
        \mid {\theta}_1=i
    \right].
\end{equation}

If the consumer accepts a contract, in each period she observes type $\theta_{t}\in\Theta$, then announces a message to the chosen firm. 
At the end of the period the income realization $y_{t}$ is observed and the customer receives (or pays) transfers from the firm as described in the chosen mechanism. 
I study (weak) Perfect Bayesian Bayesian (PBE) of this extensive form.

If the consumer's types were observed by firms, the consumer would obtain actuarily fair full insurance in equilibrium, smoothing consumption both across income realizations within a period as well as across periods. In other words, a consumer with initial type $\theta \in \left\{ l, h \right\}$ would receive a time- and income-independent flow consumption with total discounted utility

\begin{equation} \label{eq:outside-option}
    V^{FI}_{i} \equiv 
    u (c^{FI}_i)
    \sum_{t=1}^T \delta^{t-1},
\end{equation}
where $c^{FI}_i$ represents the discounted average expected lifetime income of a consumer with initial type $\theta_1 = i$:
\begin{equation*}
    c^{FI}_i
    \equiv 
    \mathbb{E}
    \left[
        \frac{
            \sum_{t=1}^T \delta^{t-1}
            {y}_t
        }{
            \sum_{t=1}^T \delta^{t-1}
        }
        \mid {\theta}_1=i
    \right].
\end{equation*}

Equilibrium outcomes are reminiscent of RS, with (initially) low-type consumers receiving their full-information utility level efficiently, while (initially) high-type consumers receive an inefficient partial coverage contract which delivers utility in the interval $\left( V^{FI}_l, V^{FI}_h \right)$. 
I assume that the full information continuation utility vector is feasible in the presence of private information, i.e., that $(V^{FI}_l,V^{FI}_h) \in \mathcal{V}$. 
This assumption insures that the following critical payoff vector, 
which will be shown to correspond to the equilibrium utility level of the consumer, is well-defined.

Define as $\Pi^*_i(V)$, for $i=l,h$, the expected discounted profit obtained by the firm conditional on the consumer's initial type being $\theta_1=i$.

\begin{lem}
    There exists a unique pair $V^* \equiv \left(  V^{*}_l, V^{*}_h \right) \in \mathcal{V}$ satisfying
    \begin{gather*}
        V^*_l = V^{FI}_l, \\
        \Pi_h \left( 
            V^*_l, V^*_h
         \right)  =0.
    \end{gather*}
\end{lem}

\begin{proof}
    Utility level $V^*_l$ can be defined as $V^{FI}_l$. The existence and uniqueness of $V^*_h$ follows from the fact that 
    $\Pi_h\left( V^*_l, \cdot  \right)$ 
    is strictly decreasing and continuous (since it is convex) and satisfies
    \begin{gather*}
        \Pi_h\left( V^*_l, V^*_l  \right) > 0,\\
        \Pi_h\left( V^*_l, V^{FI}_l  \right) < 0.
    \end{gather*}
\end{proof}

For simplicity, we also focus on equilibria with two properties.\footnote{
    The last two restrictions do not affect equilibrium outcomes but substantially simplify the analysis.
}
First, the consumer's strategy is symmetric, i.e., 
the probability she accepts the offer of firm $j$, when firms' offered mechanisms are $(M,M')$, is equal to the probability of accepting the offer of firm $j' \neq j$ when firms' offered mechanisms are exchanged. 
Second, we assume that the consumer follows the truth-telling reporting strategy whenever it is optimal.

I say that a direct mechanism $M=\left\{ z_t \right\}_{t=1}^T$ constitutes an equilibrium outcome if a PBE exists in which the on-path net consumption of the consumer corresponds exactly to their consumption in mechanism $M$ when following a truth-telling reporting strategy.
The following result --- proven in the Appendix --- shows that the unique equilibrium outcome is characterized by the solution of problem $\Pi\left( \cdot \right)$ studied in Sections \ref*{sec:Incentives_Distortions}-\ref*{sec:Utility-Distortion-dynamics}.\footnote{
    We use notation 
    $\frac{
        \partial_+ \Pi
    }{\partial V_i}
    (V)$ to denote the right-derivative of $\Pi$ with respect to $V_i$ at $V$.
}

\begin{proposition} \label{prop:Competitive-equilibrium-characterization}
    Any pure strategy PBE has outcome $M^{V^*}$.
    Moreover, a pure strategy equilibrium exists if, and only if,
    \begin{equation} \label{eq:existence condition}
        {u'\left[
            u^{-1}\left(
                c^{FI}_l
            \right)
        \right]}
        {
            \frac{\partial_{+}\Pi_{h}\left(V^{*}\right)}{\partial V_{l}}
        }
        <
        \frac{\mu_{l}}{\mu_{h}}.
    \end{equation}
\end{proposition}

The left-hand side of condition (\ref*{eq:existence condition}) does not depend on initial type distribution $(\mu_l,\mu_h)$, and hence a pure strategy equilibrium exists if, and only if, the share of (initially) low-types in the population is sufficiently large. This is in line with the classical analysis in RS, which requires that the share of high-risk consumers be sufficiently high.

Since $V^*_l<V^*_h$, 
the equilibrium outcome  in the competitive model --- described in Proposition \ref*{prop:Competitive-equilibrium-characterization} --- is the solution of a particular instance of the problem studied in Sections \ref*{sec:Incentives_Distortions}-\ref*{sec:Utility-Distortion-dynamics}.\footnote{
    The interiority of the equilibrium outcome, which is assumed in multiple characterization results, can be guaranteed as long as the income distribution induced by both types is not ``too'' distinct. 
    To see this notice that, if $p_h=p_l$, we have that $V^*_l=V^*_l$ and as a consequence the equilibrium outcome is the full information one, which is interior.
}

\subsection{The role of commitment} \label{subsec:commitment}

I have assumed so far that consumers are able to commit to a long-term mechanism.
The goal of this section is to discuss the validity of this assumption, its role in the formal analysis, and the impact of relaxing it.
If consumers have the option to renege on a mechanism at any given period and an available offer dominates their anticipated continuation contract, they would choose to do so and start a relationship with a new firm.

The assumption of consumer commitment is reasonable in the presence of institutional features that prevent or hinder consumers' contract switching.
For example, the presence of switching costs leads to lock-in effects allowing firms, at the outset, to credibly offer contracts that, at some future point in the interaction, may lead to lower continuation utility to the consumer relative to the offers available in the market 
(see \cite{honka2014quantifying} and \cite{handel2018frictions} for discussions of switching costs in auto and health insurance, respectively)

An alternative justification for commitment is informational: since consumers have private information about their types, their decision to search for a new contract may be interpreted as a negative signal by potential new firms and, as a consequence, lead to less attractive offers for switching consumers.
I now show that, if consumer's low-type is an absorbing state, firm's negative inference from consumer's switching decision may serve as a commitment device and allow for the equilibrium outcome in the model with commitment to be robust to consumer reentry.

\subsubsection*{Absorbing state}

This section assumes that the low-type is an absorbing state of the consumer type's Markov process, i.e., $\pi_{ll}=1$.
I also restrict attention to the case of realization-independent contracts.

Consider the following extension of the extensive form considered in Section \ref*{sec:competitive-analysis}.
For simplicity, assume that a different pair of firms can potentially make offers to the consumer in each period, i.e.,
the set of firms is 
$\mathcal{F} \equiv \left\{ F^j_t \mid j \in \left\{ 1 , 2 \right\}, t=1,\dots,T \right\}$.
In the first period, the consumer and firms $\left\{ F^1_1, F^2_1 \right\}$ interact as described in the baseline competitive model.\footnote{
    The assumption of two firms is purely for notational convenience.
}

However, in each period $t=2,\dots,T$ the consumer decides whether to stay with her current mechanism or reenter the market searching for a new contract. 
If the consumer decides to reenter the market, firms $F^1_t$ and $F^2_t$ observe the consumer reentry decision and then simultaneously offer a mechanism to this consumer. The consumer then decides whether to accept any of the new offers made or to remain uninsured. Regardless of period $t$'s choices, the same set of moves occurs in period $t+1$, if $t<T$.

The outcome described in Proposition \ref{prop:Competitive-equilibrium-characterization} can arise in a PBE of this extended model, as long as firms' off-path beliefs are pessimistic.
Firms' offers upon reentry in periods $t=2,\dots,T$ depends on their beliefs about the type of a consumer that decides to reenter the market. The most pessimistic belief firms may hold is to assume that this consumer has a low type in the current period for sure.
In this case, it is optimal for firms $F^1_t$ and $F^2_t$ to behave as if they were in a market without information asymmetry and offer an efficient contract which provides perfect consumption smoothing for a consumer with low-type in period $t$, i.e., with consumption flow

\begin{equation} \label{eq:outside_option}
    c^o \equiv 
    \mathbb{E} \left[ 
        \tilde{y}_t \mid \tilde{\theta}_t = l
    \right].
\end{equation}

Define $V^o_t \equiv u(c^o) \sum_{\tau=t}^T \delta^{\tau-t}$ to be the continuation utility from taking such an offer. Notice that, while extreme, these strategies are sequentially optimal for firms given their beliefs. 
Consider the strategy profile for firms such that, for $j=1,2$,
firm $F^j_1$ use the strategy described in 
the commitment model, while firm $F^j_t$ for $t\geq 2$ makes the constant consumption offer described in (\ref{eq:outside-option}).
The following lemma shows that, given firms' strategy profile, consumers never want to reenter the market.

\begin{lem}
    If the equilibrium contract $M^{V^*}$ is interior, then the consumer's continuation utility satisfies
    \begin{equation*}
        V_t(\theta^{t}) 
        \geq 
        V^o_t
    \end{equation*}
    for all periods $t=1,\dots,T$, and almost all $\theta^t \in \Theta^{t}$.
\end{lem}

\begin{proof}
    First notice that, from Proposition \ref{prop:Competitive-equilibrium-characterization} and $\pi_{ll}=1$ we have that
    $
    V_1(l) 
        =
        \sum_{\tau=1}^T \delta^{\tau-1} u(c^o).
    $

    Now consider period $t=2,\dots, T$ and history $(h^{t-1},l)$. The consumer's continuation utility in this case is given by 
    \begin{equation*}
        V_t(h^{t-1},l) 
        =
        \sum_{\tau=t}^T 
        \delta^{\tau - t} \left( 
            \nu_\tau - \Delta_\tau
        \right).
    \end{equation*}
    
    Since, from Proposition \ref{prop:RI_monotonicity}, 
    $\left\{ \nu_t \right\}_{t=1}^T$ is increasing and
    $\left\{ \nu_t \right\}_{t=1}^T$ is decreasing,
    it follows that 
    \begin{equation*}
        V_{t}(h^{t-1},l) 
        >
        V_1(l)
        \frac{
            \sum_{\tau=t+1}^T 
            \delta^{\tau - t -1}  
        }{
            \sum_{\tau=1}^T 
            \delta^{\tau - 1}    
        }, 
    \end{equation*}
    and hence the left-hand side is larger than $V_1(l)=V^o_t$.

    The only possible histories remaining are $h^t$, for $t=1,\dots,T$ and the proof is concluded since $V_t(h^t) > V_t(h^{t-1},l)$.
\end{proof}

\section{Monopoly} \label{sec:monopoly}

Now consider an interaction between a Monopolist and a consumer.
While the consumer's utility vector $(V^*_l,V^*_h)$ in the competitive model is an equilibrium object determined by firms' zero profit and no-deviation conditions, in the Monopolist problem, it is the result of an additional layer of optimization by the seller, taking into account the consumer's type-dependant participation constraint.

To be more precise, we consider the Monopolist's problem of designing a mechanism to maximize revenue, with the assumption that the consumer is privately informed about their initial type at the contracting stage. 
All future type realizations are also privately observed by the consumer.
The consumer's outside option are given by $\underline{V}_{\theta_1}$ defined in (\ref{eq:OutsideOption}).

Intuitively, the Monopolist's problem can be divided in two parts. First, for any utility level is delivered to the consumer, conditional on her initial type, the offered contract must provide these utility levels in a cost minimizing way. In other words, the optimal mechanism is the solution to $\Pi(V)$, for some $V \in \mathcal{V}$.

Second, the utility vector to be offered to the consumer, conditional on her initial type, must be chosen optimally. This gives rise to the following problem, which we denote as $\mathcal{P}^M$:
\begin{equation*}
    \max_{V \in \mathcal{V}}
    \Pi(V),
\end{equation*}
subject to 
\begin{equation*}
    V_i \geq \underline{V}_i \text{, for }i=l,h.
\end{equation*}

The next proposition shows that the Monopolist's optimal contract solves $\Pi^{*}\left(V_{l},\underline{V}_{h}\right)$, with $V_{l}<\underline{V}_{h}$. 
The participation constraint for the initially high-type buyer necessarily binds. 
However, the low-type buyer's participation constraint might be slack, i.e., the consumer may have information rents. 
This can be optimal because increasing the utility offered to the low-type buyer relaxes the binding incentive constraint in the profit maximization problem, leading to higher profits from the high-type buyers 
($\Pi_{h}\left(\cdot\right)$ increases with $V_l$).

\begin{proposition} \label{prop:monopoly-problem}
    The Monopolist's optimal offer is 
    $M^{V^M}$,
    where $V^M$ is the solution to $\mathcal{P}^M$.

    Moreover, $V^M$ satisfies:
    \begin{gather*}
        V^M_h = \underline{V}_h, \\
        V^M_l \in [\underline{V}_l, \underline{V}_h).
    \end{gather*}
\end{proposition}

\begin{proof}
    The first part of the proposition is trivial.

    We now prove that $V^M_h=\underline{V}_h$. 
    By way of contradiction, suppose this is not the case, i.e., $V^M_h > \underline{V}_h$.

    If, additionally, $V^M_l \geq V^M_h$, then mechanism $M^{V'}$, with $V' = V^M(1-\gamma) +\gamma u(0)\sum_{t=1}^T\delta^{t-1}$ for $\gamma>0$ sufficiently small is feasible and a strict improvement given concavity of $\mathcal{V}$ and $\Pi(\cdot)$.

    On the other hand, if $V^M_l < V^M_h$, then mechanism $M^{V'}$, 
    with $V' = (V^M_l,V^M_h - \varepsilon)$, for $\varepsilon>0$ sufficiently small is feasible, since it is in the convex hull of
    $\left\{ 
        V^M,
        (V^M_l,V^M_l)
    \right\} \subset \mathcal{V}$.
    It also strictly increases profits, which contradicts optimality of $V^M$.

    We now show that $V^M_l < \underline{V}_h$. If, by way of contradiction, $V^M=(\underline{V}_h , \underline{V}_h)$ then a reduction in the utility offered to the low-type is feasible and profitable since 
    \begin{equation*}
        \frac{
            \partial \Pi
        }{
            \partial V_l
        } (\underline{V}_h , \underline{V}_h)
        =
        - \mu_l \psi'(\underline{c}_h),
    \end{equation*}
    where $\underline{c}_h$ is the constant consumption flow that generates discounted utility $\underline{V}_h$ 
    (see Lemma \ref{lem:derivative-at-45degreeline} in the Appendix).


\end{proof}

Proposition \ref{prop:monopoly-problem} implies that the Monopolist's optimal mechanism is also a special instance of the problem proposed in Section \ref{Sec:profit-maximizing-contracts} and characterized in Sections \ref{sec:Incentives_Distortions}-\ref{sec:Utility-Distortion-dynamics}.
Additionally, Proposition \ref{prop:monopoly-problem} allows us to obtain a simple characterization of information rents in the optimal contract.
Given the concavity of $\Pi(\cdot)$, the utility vector 
$\underline{V} \equiv (\underline{V}_l, \underline{V}_h)$ 
solves problem $\mathcal{P}^M$ if, and only if,
\begin{equation} \label{eq:condition-binding-both-ics}
    \frac{
        \partial \Pi
    }{\partial V_l} (\underline{V})
    =
    - 
    \frac{
        \mu_l
    }{
        u' \left[ 
        u^{-1} \left( \underline{c}_l \right)
        \right]
    }
    + \mu_h
    \frac{
        \partial_+ \Pi_h
    }{
        \partial V_l
    } (\underline{V}) \leq 0,
\end{equation}
where $\underline{c}_l$ is the constant consumption flow that generates discounted utility $\underline{V}_l$.

The first term in (\ref{eq:condition-binding-both-ics}) corresponds to 
$\frac{\partial \Pi_l}{\partial V_l} (\underline{V})$,
which follows from the fact that the low-type receives an efficient continuation contract in mechanism $M^V$, for any utility vector $V$ close to $\underline{V}$. This analysis can be summarized in the following result.

\begin{proposition}
    The Monopolist's optimal mechanism leaves no information rents (i.e., $V^M=\underline{V}$) if, and only if,
    \begin{equation*}
        u' \left[ 
            u^{-1} \left( \underline{c}_l \right)
        \right]
        \frac{
            \partial_+ \Pi_h
        }{
            \partial V_l
        } (\underline{V}) 
        \leq
        \frac{
            \mu_l
        }{
            \mu_h
        }.
    \end{equation*}
\end{proposition}

In summary, the high-type receives no information rent and a distorted allocation as the least willing to pay for coverage.
On the other hand, the low-type, who is most willing to pay for coverage, receives an efficient continuation contract.
Additionally, if an initial low-type is sufficiently likely, this type receives no information rent in the optimal mechanism.
It is a key observation that the optimal mechanism \textit{always} separates different initial types into different continuation contracts, even if both participation constraints bind. 
This is a consequence of the type-dependent outside options in this model.

\section{Conclusion} \label{sec:conclusion}

This paper studies a natural dynamic extension of the workhorse insurance theoretical models proposed in the literature following \cite{rothschild1976equilibrium}, \cite{wilson1977model} and \cite{stiglitz1977monopoly}. 
The analysis introduces new tools to the dynamic mechanism design literature to deal with the presence of curvature and exploit the recursive structure of the profit maximization problem, namely the use of an auxiliary cost minimization problem and the characterization of continuation-monotonicity.

I show that the optimal contract uses both signals about consumer accidents as well as partial coverage as informative signals, affecting future offers made to consumers. 
In the case of realization-independent contracts, a strong efficiency result hods: distortions decrease along all histories. It is also shown that partial coverage contracts become more attractive over time, which implies --- if $T = \infty$ or types are fully persistent --- that the partial insurance contract offer becomes cheaper following a longer spell of partial coverage.

The assumption of time-invariance on the types and income processes can be significantly relaxed. The results in Section \ref{sec:Incentives_Distortions} can be extended to time-dependent transitions 
$\left\{ \pi^t_{ij} \right\}_{i,j , t\leq T}$ 
and income distributions 
$\left\{ p^t_i \left( \cdot \right) \right\}_{i , t\leq T }$, 
as long as types are persistent, i.e.,
$\pi^t_{ii} > \pi^t_{ji}$ for all periods.
The time-invariance of income distributions $p_i\left( \cdot \right)$ is necessary for the results in Section \ref{sec:Utility-Distortion-dynamics}, as the relevant cost-minimization problem $\mathcal{P}^A$ must be time-invariant.

The analysis of competition focuses on pure strategies.
If condition (\ref*{eq:existence condition}) does not hold, an equilibrium will necessarily involve randomization over mechanisms, following
\cite{farinha2017characterization}.
In that paper, firms randomize over the utility level offered to both types, however all offers on the equilibrium path have the property that the utility delivered to the high-type is higher that the one delivered to the low-type.\footnote{A high-type corresponds to a low-risk in the notation  used in \cite{farinha2017characterization}.}
It is a natural conjecture that a similar mixed strategy equilibrium may exist in the setting considered here, which would imply that on-path outcomes still correspond to special instances of the problem studied here.

Multiple related directions for research remain open.

The extension to multiple types leads to technical difficulties present in other dynamic mechanism design problems related to the relevance of non-local incentive constraints (see \cite{BattagliniLamba}). 
A natural conjecture that follows from my characterization is that the optimal contract features a finite menu of flow contracts with different levels of coverage, 
with reductions in coverage at a given period being rewarded, leading to subsequent menus that are more attractive.

The recursive characterization of monotonicity, in terms of flow allocation and continuation utilities, may prove useful in the study of other dynamic mechanism design problems with curvature and persistence, such as screening risk averse buyers or Mirrleesian taxation problems.

A crucial question in dynamic contracting is the issue of commitment, especially on the side of consumers. The discussion in Subsection \ref{subsec:commitment} illustrates the role that information asymmetry across firms may play as a commitment device. However, it is important to understand how the optimal contract structure changes as we consider more complex models where consumer reentry may be driven by preference or exogenous separation shocks leading to presence of switching consumers with high-types, improving firms' beliefs about switching consumers and hence leading to more attractive outside offers.

\bibliographystyle{ecta}
\bibliography{databaseVitorLink}

\pagebreak

\section*{Appendix} \label{Sec:Appendix}

\setcounter{section}{0}

\section{Recursive formulation} 

In this section, I provide a recursive representation of the firm's relaxed problem described in (\ref{eq:RelaxedProblem})
This representation is then used to prove Lemma \ref{Lem:Properties_relaxed_problem}.
To explore the recursive structure of the problem at hand, we define, for any period, the set of feasible continuation utilities that can be delivered to a consumer for any given current type as well as the maximal continuation profit that can be obtained by the firm. 

\subsection{Definitions}

\subsubsection{Utility mechanisms}

For tractability, I will represent direct mechanisms through the
utility flow generated for each period and history. Consider any direct
mechanism $M=\left\{ z_{t}\right\} _{t=1}^{T}$, period $t$ and sequence
of types and incomes 
$\left(\theta^{t},y^{t}\right) \in \Theta^t  \times Y^t$. 
The utility flow generated in period $t$ is $\vartheta_{t}\left(\theta^{t},y^{t}\right)=u\left(z_{t}\left(y_{t}\mid\eta^{t-1},\theta_{t}\right)\right)$,
where $\eta^{t-1}=\left(\theta^{t-1},\left(\phi\left(y_{1}\right),...,\phi\left(y_{t-1}\right)\right)\right)$.
I define a utility-direct mechanism (UDM) as a sequence of functions
$M_{1}^{u}\equiv\left\{ \vartheta_{t}\right\} _{t=1}^{T}$, with $\vartheta_{t}:\Theta^{t}\times Y^{t}\mapsto u\left(\mathbb{R}\right)$,
such that $\vartheta_{t}\left(\theta^{t},y^{t}\right)$ is $\left(\eta^{t-1},\theta_{t},y_{t}\right)$-measurable.
In other words, we impose that the utility flow in period $t$ only
depend on the history of incomes $y^{t-1}$ through the observable signal history
$\phi^{t-1}$. 
Similarly, a period $t$ continuation
UDM assigns utility flows beyond period $t$ which depend on the history
of types and income levels starting at period $t$, i.e., it is equal
to $M_{t}^{u}\equiv\left\{ \vartheta_{\tau}\right\} _{\tau=t}^{T}$ such
that $\vartheta_{\tau}:\Theta^{\tau-t}\times Y^{\tau-t}\mapsto u\left(\mathbb{R}\right)$
and $\vartheta_{\tau}\left(\left[\theta^{\tau}\right]_{t}^{\tau},\left[y^{\tau}\right]_{t}^{\tau}\right)$
is $\left(\left[\eta^{\tau}\right]_{t}^{\tau-1},\theta_{\tau} , y_\tau\right)$-measurable.
I define the set of period $t$ continuation UDM such that no one-shot
OSCD with a misreport in periods $\tau\geq t$ is profitable as
$\mathcal{M}_{t,OSIC}^{u}$.

\subsubsection{Profits}

Define, for any $t=1,...,T$, the set of type-contingent continuation utilities 
generated by incentive compatible continuation mechanisms:
\[
\mathcal{V}_{t}\equiv\left\{ \left(V_{l},V_{h}\right)\in\mathbb{R}^{2}
\mid
\begin{array}{c}
\exists M_{t}^{u}\equiv\left\{ \vartheta_{\tau}\right\} _{\tau=t}^{T}\in\mathcal{M}_{t,OSIC}^{u}\text{ s.t.}\\
V_{i}
=
\mathbb{E}\left[\sum_{\tau=t}^{T}\delta^{\tau-t}\vartheta_{\tau}\left(\left[\theta^{\tau}\right]_{t}^{\tau},\left[y^{\tau}\right]_{t}^{\tau}\right)\mid\theta_{t}=i\right]\text{, for }i=l,h.
\end{array}\right\} ,
\]
with $\mathcal{V}_{T+1}\equiv\left\{ \left(0,0\right)\right\} $ if
$T<\infty$. 
It is easy to show that these sets are convex. For any
$V\in\mathcal{V}_{t}$, define the problem $\mathcal{P}_{t}\left(V\right)$
of finding the maximal continuation profit obtained by a firm, conditional on continuation utility levels $V$, as
\begin{equation}
    \Pi_{t}\left(V\right)
    \equiv
    \sup_{M_{t}^{u}\in\mathcal{M}_{t,OSIC}^{u}}
    \mathbb{E}\left[
        \sum_{\tau=t}^{T} \delta^{\tau-t}\left[y_{t}-\psi\left[\vartheta_{\tau}\left(\left[\theta^\tau\right]_{t}^{\tau},\left[y^\tau\right]_{t}^{\tau}\right)\right]\right]\mid\theta^{t-1}=h^{t-1}
    \right],
    \label{eq:profit_max-1}
\end{equation}
subject to, for $i\in\left\{ l,h\right\}$,
\begin{equation}
    V_{i}=\mathbb{E}\left[\sum_{\tau=t}^{T}\delta^{\tau-t}
    \vartheta_{\tau}\left(\left[\theta^\tau\right]_{t}^{\tau},\left[y^\tau\right]_{t}^{\tau}\right)
    \mid
    \theta^{t}=\left(h^{t-1},i\right)\right],\label{eq:PK}
\end{equation}
with $\Pi_{T+1}\left(0,0\right)=0$ if $T<\infty$. Denote
as $M_{t}^{*,V}\in\mathcal{M}_{t,OSIC}^{u}$ the solution to this
problem, whenever is exists and is unique.

Finally, for $v\in\frac{1-\delta^{T-t+1}}{1-\delta}u\left(\mathbb{R_{+}}\right)$,
define the full information continuation profit function as 
\[
\Pi_{t,i}^{FI}\left(v\right)\equiv\sum_{\tau=t}^{T}\delta^{\tau-t}\left[\mathbb{E}\left[{y}_{t}\mid\theta_{t}=i\right]-\psi\left(\frac{1-\delta}{1-\delta^{T-t+1}}v\right)\right].
\]

It is easy to show that $\Pi_{t}\left(V\right)\leq\pi_{hl}\Pi_{t,l}^{FI}\left(V_{l}\right)+\pi_{hh}\Pi_{t,h}^{FI}\left(V_{h}\right)$.
If $V_{l}\geq V_{h}$, both functions coincide since offering constant
utility $\frac{1-\delta}{1-\delta^{T-t+1}}V_{i}$, following type
$\theta_{t}=i$, is feasible (incentive constraints of relaxed problem
do not bind). 

\subsection{Recursive representation}

The following preliminary result is useful in studying the recursive structure of the relaxed problem.
\begin{lem} \label{lem:uniqueness-concavity}
If problem $\Pi_{t}\left(\cdot\right)$ has a solution then:

(i) It is strictly concave,

(ii) Its solution is unique.
\end{lem}
\begin{proof}

Proof of (i). For any $t$, consider utility pairs $V^{1},V^{2}\in\mathcal{V}_{t}$,
with $V^{1}\neq V^{2}$, and $\alpha\in\left(0,1\right)$. 
For $k=1,2$, take optimal period $t$ continuation UDMs 
$M_{t}^{u,k}\equiv\left\{ \vartheta_{\tau}^{k}\right\} _{\tau=t}^{T}$ 
in problem $\Pi_{t}\left(V^{i}\right)$. The mechanism $M_{t}^{u,\alpha}=\left\{ \alpha\vartheta_{\tau}^{1}+\left(1-\alpha\right)\vartheta_{\tau}^{2}\right\} _{\tau=t}^{T}$
is feasible in $\Pi_{t}\left(V^{\alpha}\right)$ and, since
the objective function in (\ref{eq:profit_max-1}) is strictly concave,
generates profits strictly above $\alpha\Pi_{t}\left(V^{1}\right)+\left(1-\alpha\right)\Pi_{t}\left(V^{2}\right)$.
Since all incentive constraints as well as (\ref{eq:PK}) are linear
in utility flows, we have that $M_{t}^{u,\alpha}\equiv\left\{ \vartheta_{\tau}^{k}\right\} _{\tau=t}^{T}$
is feasible in $\Pi_{t}\left(V^{\alpha}\right)$. Hence $\Pi_{t}\left(V^{\alpha}\right)>\alpha\Pi_{t}\left(V^{1}\right)+\left(1-\alpha\right)\Pi_{t}\left(V^{2}\right)$.

Proof of (ii). Follows similarly from strict concavity of objective
function and linearity of constraints in problem $\Pi_{t}\left(\cdot\right)$.
\end{proof}

We extend the definition of $\xi$ by defining, for any function $\vartheta:Y\mapsto u\left(\mathbb{R}_{+}\right)$,
\[
\xi\left(\vartheta,\theta\right)\equiv\sum_{y\in Y}p_{\theta}\left(y\right)\left[y-\psi\left[\vartheta\left(y\right)\right]\right].
\]

Define a period $t$ policy as a any pair $\left(\vartheta,N\right)$ such
that $\vartheta:Y\mapsto u\left(\mathbb{R}_{+}\right)$ and $N:\Phi\mapsto\mathcal{V}_{t+1}$,
and the set of period $t$ policies as $\mathcal{N}_{t}$. Finally,
define 
\[
\Gamma_{t}\equiv\left\{ \left(V_{l},V_{h}\right)\mid\begin{array}{c}
\exists\left(\vartheta,N\right)\in\mathcal{N}_{t}\text{ such that}\\
V_{h}=\sum_{y\in Y}p_{h}\left(y\right)\left[\vartheta\left(y\right)+\pi_{hh}N_{h}\left(\phi\left(y\right)\right)+\pi_{hl}N_{l}\left(\phi\left(y\right)\right)\right]\\
V_{l}\geq\sum_{y\in Y}p_{l}\left(y\right)\left[\vartheta\left(y\right)+\pi_{hh}N_{h}\left(\phi\left(y\right)\right)+\pi_{hl}N_{l}\left(\phi\left(y\right)\right)\right]\\
V_{l}\in\frac{1-\delta^{T-t+1}}{1-\delta}u\left(\mathbb{R}_{+}\right)
\end{array}\right\} ,
\]
and the following optimization problem, choosing the optimal period
$t$ policy:
\begin{equation}
P_{t}\left(V\right)\equiv\begin{array}{c}
\sup_{\left(\vartheta,N\right)\in\mathcal{N}_{t}}\pi_{hh}\left[\xi\left(\vartheta,h\right)+\delta\sum_{y\in Y}p_{h}\left(y\right)\Pi_{t+1}\left(N\left(\phi\left(y\right)\right)\right)\right]+\pi_{hl}\Pi_{t,l}^{FI}\left(V_{l}\right)\\
\text{s.t. }\left\{ \begin{array}{c}
V_{l}\geq\sum_{y\in Y}p_{l}\left(y\right)\left\{ \vartheta\left(y\right)+\delta\left[\pi_{lh}N_{h}\left(\phi\left(y\right)\right)+\pi_{ll}N_{l}\left(\phi\left(y\right)\right)\right]\right\} \\
V_{h}=\sum_{y\in Y}p_{h}\left(y\right)\left\{ \vartheta\left(y\right)+\delta\left[\pi_{hh}N_{h}\left(\phi\left(y\right)\right)+\pi_{hl}N_{l}\left(\phi\left(y\right)\right)\right]\right\} 
\end{array}\right.
\end{array},\label{eq:recusrive_profit}
\end{equation}
with solution, whenever it exists, denoted as $\left(\vartheta_{t}^{V},N_{t}^{V}\right)$.

\begin{lem} \label{lem:show-recursivity} 
    For any $t=1,...,T$ and $V\in\mathcal{V}_{t}$, let $\left\{ \vartheta_{\tau}^{*}\right\} _{\tau=t}^{T}$ be the solution to problem (\ref{eq:profit_max-1}). 
    The following hold, for any $t\leq T$:

    (i) 
    (Full insurance following low type)
    if UDM 
    $\left\{ \vartheta_{t}^{*}\right\} _{\tau = t}^T$ solves  $\Pi_t(V)$,
    then, for any 
    $1 \leq \tau \leq T-t+1$
    and
    $y^{\tau -1} \in Y^{\tau-1}$,
    \begin{equation*}
        \vartheta_{t + \tau'}(\theta^{\tau'},y^{\tau'})
        =
        \vartheta_{t + \tau''}(\theta^{\tau''},y^{\tau''})
    \end{equation*}
    if $\theta^{\tau'} , \theta^{\tau''} \succeq (h^{\tau-1},l)$ and
    $y^{\tau'} , y^{\tau''} \succeq y^{\tau -1 }$.\footnote{
        For completeness, we define $Y^0 \equiv \left\{ \emptyset \right\}$ 
        and impose that all income histories succeed $\emptyset$.
    }



    (ii) Recursive utility set representation: $\mathcal{V}_{t}=\Gamma_{t}$,

    (iii) $\Pi_{t}\left(\cdot\right)$ satisfies the following recursive
    equation, i.e., $\Pi_{t}\left(\cdot\right)=P_{t}\left(\cdot\right)$,

    (iv) Flow contract and utilities following high type: 
    for any $y\in Y$, $V \in \mathcal{V}_t$ and $i\in\left\{ l,h\right\} $,
    \[
    \vartheta_{t}^{*}\left(h,y\right)=\vartheta_{t}^{V}\left(y\right),
    \]
    \[
    \mathbb{E}\left[\sum_{\tau=t+1}^{T}\delta^{\tau-t-1}\vartheta\left(\left[\theta\right]_{t}^{\tau},\left[y\right]_{t}^{\tau}\right)\mid\theta^{t}=\left(h^{t},i\right),y_{t}=y\right]=N_{i}^{V}\left(\phi\left(y\right)\right).
    \]
\end{lem}

\begin{proof}
We start by proving (i).

Consider problem $\Pi_t(V)$, 
$1 \leq \tau \leq T-t+1$,
 and $y^{\tau -1 } \in Y^{\tau-1}$.
Utility flows for histories 
$(\theta^{\tau'}, y^{\tau'})$ satisfying 
$\theta^{\tau'} \succeq (h^{\tau-1},l)$ and
$y^{\tau'} \succeq y^{\tau -1 }$
only affect the objective function through
\begin{equation} \label{eq:loss_follow_l}
    \mathbb{E}\left\{
        \sum_{\tau' = t + \tau -1}^{T}
        \delta^{\tau'-t }\left[
            y_{\tau'} - 
            \psi\left[
                \vartheta_{\tau'}\left(\left[\theta^{\tau'} \right]_{t}^{\tau'},\left[y^{\tau'}\right]_{t}^{\tau'}\right)
            \right]
        \right]
        \mid
        [\theta^T]_t^{t+\tau-1}=(h^{\tau -1},l),
        [y^T]_t^{t+\tau-2}=y^{\tau-1}
    \right\},
\end{equation}

and they only affect incentive constraints through 

\begin{equation} \label{eq:Vl_promise}
    \mathbb{E}\left\{
        \sum_{\tau' = t + \tau -1}^{T}
        \delta^{\tau'-t }
        \vartheta_{\tau'}\left(\left[\theta^{\tau'} \right]_{t}^{\tau'},\left[y^{\tau'}\right]_{t}^{\tau'}\right)
        \mid
        [\theta^T]_t^{t+\tau-1}=(h^{\tau -1},l),
        [y^T]_t^{t+\tau-2}=y^{\tau-1}
    \right\},
\end{equation}

Convexity of $\psi$ implies that a constant utility flow, following
type realizations $(h^{\tau -1},l)$
and income realizations 
$y^{\tau-1}$
uniquely maximizes
(\ref{eq:loss_follow_l}), subject to (\ref{eq:Vl_promise}). 
Hence any solution to the profit maximization problem $\Pi_t$ must satisfy (i),
in which case (\ref{eq:loss_follow_l}) is equal to $\Pi_{t,l}^{FI}\left(V_{l}\right)$.

Hence, we can restrict the choice set in the profit minimization problem to the set 
$\bar{\mathcal{M}}_{t,OSIC}^{u}$, which is the subset of 
${\mathcal{M}}_{t,OSIC}^{u}$ satisfying (i). As a consequence, the continuation mechanism following a low-type announcement is pinned down by the continuation utility of the consumer: it corresponds to the constant consumption amount that delivers such continuation utility.

We now show (ii)-(iv).

Fix a period any $t$. 
Any OSIC mechanism in $\bar{\mathcal{M}}_{t,OSIC}^{u}$ can be described by the continuation utility obtained conditional on $\theta_{t}=l$, the within-period $t$ utility flows generated to a consumer with $\theta_{t}=h$ and the continuation utility flows provided from period $t+1$ onwards for a consumer with $\theta_{t}=h$ (which constitutes a period $t+1$
OSIC mechanism). This statement is formalized below.

Define 
$\bar{A}_{t}\equiv\mathbb{R}\times\left[u\left(\mathbb{R}\right)\right]^{Y}\times\left[\mathcal{M}_{t+1,OSIC}^{u}\right]^{\Phi}$
as the set of triples 
$(V_l, \vartheta(\cdot), \bar{N} (\cdot))$
where, in period $t$ with type announcement $h$ and income realization $y \in Y$, $\vartheta(y)$ represents flow utility in period $t$ and
$\bar{N} (\phi(y)) \in \mathcal{M}_{t+1,OSIC}^{u}$
represents the continuation mechanism offered from period $t+1$ onwards.
The flow utility in a continuation mechanism 
$\bar{N} (\phi(y))$,
if history $(\theta^\tau, y^\tau)$ are observed starting in period $t+1$
is denoted as $\bar{N}\left(\theta^{\tau},y^{\tau} ; \phi\left(y\right)\right)$. 
So a one-to-one mapping exists between the set $\bar{\mathcal{M}}_{t,OSIC}^{u}$ and the set $A_{t}$ defined as
\begin{equation}
\left\{ 
    \left(V_{l},\vartheta\left(\cdot\right),\bar{N}\left(\cdot\right)\right)\in\bar{A}_{t}
    \mid
    \begin{array}{c}
    \exists \left(V_{l}',V_{h}'\right)\in\mathbb{R}^{2}\text{ s.t.}\\
    V_{i}'\left(\phi\right)=\\
    \mathbb{E}\left[
        \sum_{\tau=t+1}^{T}\delta^{\tau-t-1}\bar{N}\left(\left[\theta^{\tau},y^{\tau}\right]_{t+1}^{\tau};\phi\right)\mid\theta_{t+1}=i
    \right],\\
    V_{l} \geq 
    \sum_{y\in Y}p_{h}\left(y\right)\left[
        \vartheta\left(y\right)
        + \delta 
        (\pi_{lh}V_{h}'\left(\phi\left(y\right)\right)+\pi_{ll}V_{l}'\left(\phi\left(y\right)\right))
    \right],\\
    V_{l}\in\frac{1-\delta^{T-t+1}}{1-\delta}u\left(\mathbb{R}_{+}\right).
    \end{array}
\right\} .\label{eq:aux_setA}
\end{equation}

The one-to-one mapping $a:A_{t}\mapsto\bar{\mathcal{M}}_{t,OSIC}^{u}$
assigns, for each $\left(V_{l},\vartheta\left(\cdot\right),\bar{N}\left(\cdot\right)\right)\in A_{t}$,
the mechanism $a\left(V_{l},\vartheta\left(\cdot\right),\bar{N}\left(\cdot\right)\right)=\left\{ \vartheta_{\tau}\right\} _{\tau=t}^{T}$
satisfying: $\vartheta_{\tau}\left(\theta^{\tau-t},y^{\tau-t}\right)=\frac{1-\delta}{1-\delta^{T-t+1}}V_{l}$,
for any $\left(\theta^{\tau-t},y^{\tau-t}\right)\succeq\left(l,y\right)$
and any $y\in Y$; $\vartheta_{t}\left(h,y\right)=\vartheta\left(y\right)$ for
any $y\in Y$; and for any $\tau\geq t+1$ and $\left(\theta^{\tau-t-1},y^{\tau-t-1}\right)$,
$\vartheta_{\tau}\left(\left(h,\theta^{\tau-1}\right),\left(y,y^{\tau-1}\right)\right)=\bar{N}\left(\theta^{\tau-1},y^{\tau-1}\mid\phi\left(y\right)\right)$.
The mechanism $a\left(V_{l},\vartheta\left(\cdot\right),\bar{N}\left(\cdot\right)\right)$
satisfies OSIC since: (i) the inequality in expression (\ref{eq:aux_setA})
implies that the period $t$ one-shot incentive constraint is satisfied,
and (ii) the one-shot incentive constraints for periods $\tau\geq t+1$
are guaranteed since the continuation mechanism $\bar{N}\left(\phi\left(y_{t}\right)\right)$,
for any $y_{t}\in Y$, is contained in $\mathcal{M}_{t+1,OSIC}^{u}$.
It is easy to show that, for any $M_{t}^{u}\in\bar{\mathcal{M}}_{t,OSIC}^{u}$,
the vector $a^{-1}\left(M_{t}^{u}\right)$ is an element of $A_{t}$.
Moreover, the set of utilities generated by mechanisms $a\left(V_{l},\vartheta\left(\cdot\right),\bar{N}\left(\cdot\right)\right)$,
for all $\left(V_{l},\vartheta\left(\cdot\right),\bar{N}\left(\cdot\right)\right)\in A_{t}$
coincides with $\Gamma_{t}$, which implies (ii).

Now consider any $V\in\mathcal{V}_{t}$ and any mechanism $a\left(V_{l},\vartheta\left(\cdot\right),\bar{N}\left(\cdot\right)\right)\in\bar{\mathcal{M}}_{t,OSIC}^{u}$
feasible in the problem defining $\Pi_{t}\left(V\right)$. Define
$N=\left(N_{l},N_{h}\right):\Phi\mapsto\mathbb{R}^{2}$, for each
$i\in\left\{ l,h\right\} $, by
\[
    N_{i}\left(\phi\right)
    \equiv
    \mathbb{E}\left[
        \sum_{\tau=t+1}^{T}\delta^{\tau-t-1}\bar{N}\left(\left[\theta^{\tau},y^{\tau}\right]_{t+1}^{\tau};\phi\right)\mid\theta_{t+1}=i
    \right].
\]

The new mechanism $a\left(V_{l},\vartheta\left(\cdot\right),M_{t+1}^{*,N\left(\cdot\right)}\right)$,
where $M_{t+1}^{*,N\left(\phi\right)}\in\mathcal{M}_{t+1,OSIC}^{u}$
is the optimal mechanism in the problem $\mathcal{P}_{t+1}\left(N_{l}\left(\phi\right),N_{h}\left(\phi\right)\right)$,
for all $\phi\in\Phi$, is also in $A_{t}$, generates the same continuation
utility in period $t$, for any $\theta_{t}\in\left\{ l,h\right\} $,
and generates strictly higher profits if $N_{t+1}^{*,N\left(\cdot\right)}\neq\bar{N}$.
Hence, without loss of optimality in problem $\mathcal{P}_{t}\left(V\right)$,
we can focus on mechanisms indexed by utility level $V_{l}$ and mappings
$\left(\vartheta,N\right):Y\mapsto u\left(\mathbb{R}_{+}\right)\times\mathcal{V}_{t+1}$,
which are given by $a\left(V_{l},\vartheta\left(\cdot\right),M_{t+1}^{*,N\left(\cdot\right)}\right)$.
From now on, we refer to such mechanisms via the triple $\left(V_{l},\vartheta,N\right)\in u\left(\mathbb{R}_{+}\right)\times\left[u\left(\mathbb{R}_{+}\right)\right]^{Y}\times\left[\mathcal{V}_{t+1}\right]^{\Phi}$.
The mechanism connected with $\left(V_{l},\vartheta,N\right)$ satisfies
OSIC and constraint (\ref{eq:PK}) if, and only if, it satisfies
\begin{equation}
    V_{l}
    \geq
    \sum_{y\in Y} p_{l} \left(y\right) \left\{
        \vartheta\left(y\right)
        + \delta \left[ 
            \pi_{hh}N_{h}\left(\phi\left(y\right)\right)+\pi_{hl}N_{l}\left(\phi\left(y\right)\right)
        \right]
    \right\},
    \label{eq:recursive_IC}
\end{equation}
\[
V_{h}=\sum_{y\in Y}p_{h}\left(y\right)\left\{ \vartheta\left(y\right)+\delta\left[\pi_{hh}N_{h}\left(\phi\left(y\right)\right)+\pi_{hl}N_{l}\left(\phi\left(y\right)\right)\right]\right\} 
\]
 and it generates profits
\[
    \pi_{hh}\left[
        \xi\left(\vartheta,h\right)
        + \delta \sum_{y\in Y}p_{h}\left(y\right)\Pi_{t+1}\left(N\left(\phi\left(y\right)\right)\right)
    \right]
    +\pi_{hl}\Pi_{t,l}^{FI}\left(V_{l}\right).
\]
Hence, the maximal profit $\Pi_{t}\left(\cdot\right)$ must satisfy
(\ref{eq:recusrive_profit}) and the profit maximizing continuation
UDM must be generated by the solution to (\ref{eq:recusrive_profit}).
This implies (iii) and (iv).
\end{proof}

\subsection{Properties}

We use notation $\partial_{i}^{+}\Pi_{t}\left(\cdot\right)$ and $\partial_{i}^{-}\Pi_{t}\left(\cdot\right)$
to represent the right- and left-derivatives of $\Pi_{t}\left(\cdot\right)$
with respect to $V_{i}$, for $i=l,h$.
For any time $t$, continuation utility vector $V \in int (\mathcal{V}_{t})$ and constant $c$, we denote the expression
\[
    0 \in [\partial_{i}^+ \Pi_{t}(V) + c, \partial_{i}^- \Pi_{t}(V) + c]
\]
simply by 
$ 0 \in \partial_i \Pi_{t}( V ) + c.$

\begin{lem} \label{lem:derivative-at-45degreeline}
    For any $V\in int (\mathcal{V}_{t})$ with $V_l \geq V_h$,
    $\Pi_t (\cdot)$ is differentiable at $V$ and
    \begin{equation*}
        \frac{\partial}{\partial V_{i}}\Pi_{t}\left(V\right)
        =
        \pi_{hi} \frac{d}{dV_i} \Pi_{t,i}^{FI}\left(V_{i}\right)
        =
        \pi_{hi} \psi'(u_i), 
    \end{equation*}
    where $ \left( \sum_{\tau=t}^T \delta^{\tau - t} \right) u_i = V_i$
\end{lem}

\begin{proof}

    Since $\Pi_{t}\left(V\right)=\pi_{hh}\Pi_{t,h}^{FI}\left(V_{h}\right)+\pi_{hl}\Pi_{t,l}^{FI}\left(V_{l}\right)$,
    for $V_{l}\geq V_{h}$, the result is obvious for $V_{l}>V_{h}$.
    Now consider a pair $\left(V_{0},V_{0}\right)\in\mathcal{V}_{t}$,
    with $V_{0}\equiv\frac{1-\delta^{T-t+1}}{1-\delta}u_{0}$ and $u_{0}>u\left(0\right)$.
    The optimal period $t$ policy $\left(\vartheta^{V},N^{V}\right)$ induces
    constant utility flow $u_{0}$, i.e., $\vartheta^{V}\left(y\right)=u_{0}$,
    for all $y\in Y$. Now for $\varepsilon$ sufficiently small, define
    an alternative policy $\left(\tilde{\vartheta},N^{V}\right)$ with utility
    flow given by 
    \[
        \tilde{\vartheta}^{\varepsilon}\left(y\right)
        \equiv 
        u_{0} 
        + \varepsilon \frac{
            p_{l}\left(y\right)^{-1}- |Y|
        }{
            \sum_{y\in Y}\ell\left(y\right)^{-1} - |Y|
        }.
    \]
    This alternative policy is feasible in problem $\mathcal{P}_{t}\left(V_{0}+\varepsilon,V_{0}\right)$,
    which implies
    \[
    \Pi_{t}\left(V_{0}+\varepsilon,V_{0}\right)-\Pi_{t}\left(V_{0},V_{0}\right)\geq\pi_{hh}\sum_{y\in Y}p_{h}\left(y\right)\left\{ \psi\left(u_{0}\right)-\psi\left[\tilde{\vartheta}^{\varepsilon}\left(y\right)\right]\right\} ,
    \]
    holding as an equality for $\varepsilon=0$. Since the right-hand
    side (RHS) is differentiable and the left-hand side is strictly concave,
    their derivative coincides. The derivative of the RHS at $\varepsilon=0$
    is $\pi_{hh}\psi'\left(u_{0}\right)=\pi_{hh}\frac{d}{dv}\Pi_{t,h}^{FI}\left(V_{0}\right)$,
    which pins down $\frac{\partial}{\partial V_{h}}\Pi_{t}\left(V_{0},V_{0}\right)$.
    A similar argument can be used to find $\frac{\partial}{\partial V_{l}}\Pi_{t}\left(V_{0},V_{0}\right)$,
    with an $\varepsilon$-perturbation that is feasible in problem $\mathcal{P}_{t}\left(V_{0},V_{0}+\varepsilon\right)$,
    for $\varepsilon$ small.     
\end{proof}

The following result shows what type of distortions arise in the profit
maximizing mechanism and the dynamic behavior of continuation utilities.

\begin{lem} \label{lem:recursive-properties}
    For any $V\in int\left(\mathcal{V}_{t}\right)$, the solution 
    $\left(\vartheta^{V},N^{V}\right)$ 
    of (\ref{eq:recusrive_profit})
    satisfies: there exists multipliers $\lambda \geq 0$ and $\mu > 0$ such
    that:

    (i) Multiplier signal: $\lambda>0$ and both constraints in 
    (\ref{eq:recusrive_profit})
    hold as equalities if, and only if, $V_{h}>V_{l}$,

    (ii) Within-period distortions: $\vartheta_{t}^{V}$ satisfies 
    \[
    -\psi'\left(\vartheta^{V}\left(y\right)\right)+\mu-\lambda\ell\left(y\right)\begin{cases}
    \leq0 & \vartheta^{V}\left(y\right)=u\left(0\right),\\
    =0 & \text{, if }\vartheta^{V}\left(y\right)>u\left(0\right).
    \end{cases}
    \]

    (iii) Continuation utility rewards high types: if $V_{h}>V_{l}$, $t<T$,
    then 
    \[
    N_{h}^{V}\left(\phi\right)>N_{l}^{V}\left(\phi\right).
    \]
    for any $\phi \in \Phi$ such that 
    $
        N_{l}^{V}\left(\phi\right),
        N_{h}^{V}\left(\phi\right)
        >\frac{1-\delta^{T-t+1}}{1-\delta}u_{0}
    $,

    (iv) Continuation utility signal rewards: 
    for any $\phi, \phi' \in \Phi$
    \[
        \ell (\phi') \leq \ell (\phi) \implies
        \pi_{ll} N_{l}^{V}\left(\phi'\right) + \pi_{lh} N_{h}^{V}\left(\phi'\right)
        \geq 
        \pi_{ll} N_{l}^{V}\left(\phi\right) + \pi_{lh} N_{h}^{V}\left(\phi\right),
    \]
    with this inequality holding strictly if $V_h > V_l$ and $N^ {V}(\phi) \in int(\mathcal{V}_{t+1})$.
\end{lem}

\begin{proof}
Consider arbitrary period $t$ and $V\in int\left(\mathcal{V}_{t}\right)$.

Step 1. There exists period t policy $\left(\vartheta,N\right)\in\mathcal{N}_{t}$ feasible in problem $\mathcal{P}_{t}\left(V\right)$ such that inequality constraint in (\ref{eq:recusrive_profit}) holds strictly:
just consider a feasible policy in problem $\mathcal{P}_{t}\left(V_{l}-\varepsilon,V_{h}\right)$, for $\varepsilon>0$ small.

Step 2. The optimization problem $\mathcal{P}_{t}\left(V\right)$
has a concave objective, convex choice set $\mathcal{N}_{t}$ and
linear constraints. Step 1 implies that  a feasible policy $\left(\vartheta^{V},N^{V}\right)\in\mathcal{N}_{t}$
for problem $\mathcal{P}_{t}\left(V\right)$ is optimal if, and only
if, there exists multipliers $\lambda \geq 0$ and $\mu\in\mathbb{R}$
such that $\left[\left(\vartheta^{V},N^{V}\right),\left(\lambda\right)\right]\in\mathcal{N}_{t}\times\mathbb{R}_{+}$
form a saddle point of the Lagrangian
\begin{multline}
    \xi\left(\vartheta,h\right)+\sum_{y\in Y}p_{h}\left(y\right)\vartheta\left(y\right)\left[\mu-\lambda\ell\left(y\right)\right]
    + \lambda V_l - \mu V_h\\
    + 
    \sum_{\phi \in \Phi} p_{h}\left( \phi \right)
    \left\{ \begin{array}{c}
        \Pi_{t+1}\left(N\left(\phi\left(y\right)\right)\right)+\left[\mu\pi_{hh}-\lambda\ell\left( \phi \right)\pi_{lh}\right]N_{h}\left(\phi\left(y\right)\right)\\
        +\left[\mu\pi_{hl}-\lambda\ell\left( \phi \right)\pi_{ll}\right]N_{l}\left(\phi\left(y\right)\right)
        \end{array}
        \right\}.
        \label{eq:Lagrangian}
\end{multline}

Moreover, the necessary condition for optimality of $N(\cdot)$ is 
\begin{equation}
    0 \in \partial_i \Pi_{t+1}(N ( \phi (y))) + \mu \pi_{hi} - \lambda \ell(\phi) \pi_{li}, \label{FOC_N}
\end{equation}

while (ii) is a necessary condition for optimality of $\vartheta(\cdot)$. 
Since $\partial_h^{-} \Pi_{t+1}(N ( \phi (y))) < 0$, (\ref*{FOC_N}) implies $\mu > 0$.

Step 3 (item i). If $\lambda=0$, (\ref{eq:Lagrangian}) becomes the
Lagrangian of the problem $\bar{\mathcal{P}}_{t}\left(V\right)$,
where the period $t$ incentive constraint is ignored, which has as
unique solution constant flow utility equal to $\frac{1-\delta}{1-\delta^{T-t+1}}V_{i}$,
following announcement $\theta=i$, which only satisfies the period
$t$ one-shot incentive constraint if $V_{h}\leq V_{l}$. Alternatively,
if $V_{l}\geq V_{h}$, the optimal mechanism in $\mathcal{P}_{t}\left(V\right)$
involves constant utilities that do not depend on income $y$, which
is only optimal in (\ref{eq:Lagrangian}) if $\lambda=0$.

Step 4 (item ii). Property (ii) is equivalent to local optimality of $\vartheta^{V}$
in (\ref{eq:Lagrangian}).

Step 6 (item iii). 
If $V_{l}<V_{h}$ and $N_{h}^{V}\left(\phi_{0}\right)\leq N_{l}^{V}\left(\phi_{0}\right)$
for some $\phi_{0}\in\Phi$, a contradiction follows. Necessary condition (\ref{FOC_N}) implies
\[
\partial_{h}^{+}\Pi_{t+1}\left(N^{V}\left(\phi_{0}\right)\right)+\left[\mu\pi_{hh}-\lambda\ell\left(\phi_{0}\right)\pi_{lh}\right]
\leq
0
\leq
\partial_{l}^{-}\Pi_{t+1}\left(N^{V}\left(\phi_{0}\right)\right)+\left[\mu\pi_{hl}-\lambda\ell\left(\phi_{0}\right)\pi_{ll}\right]
\]
which, using $\lambda>0$, gives us
\begin{equation} \label{eq:Diff_N_compare}
    \frac{\partial_{h}^{+}\Pi_{t+1}\left(N^{V}\left(\phi_{0}\right)\right)}{\pi_{hh}}
    \leq
    \lambda\ell\left(\phi_{0}\right)\frac{\pi_{lh}}{\pi_{hh}}-\mu
    <
    \lambda\ell\left(\phi_{0}\right)\frac{\pi_{ll}}{\pi_{hl}}-\mu
    \leq
    \frac{\partial_{l}^{-}\Pi_{t+1}\left(N^{V}\left(\phi_{0}\right)\right)}{\pi_{hl}}.
\end{equation}

Finally, $N_{h}^{V}\left(\phi_{0}\right)\leq N_{l}^{V}\left(\phi_{0}\right)$ and Lemma \ref{lem:derivative-at-45degreeline} imply that $\Pi_{t+1}$ is differentiable at $V$ and

\[ 
    \frac{\partial_{i}\Pi_{t+1}\left(N^{V}\left(\phi_{0}\right)\right)}{\pi_{hi}}
    =
    \frac{d}{dV_{i}} \Pi_{t+1,i}^{FI}\left(N_{i}^{V}\left(\phi_{0}\right)\right)
    =
    \frac{d}{dV_{l}} \Pi_{t+1,l}^{FI}\left(N_{i}^{V}\left(\phi_{0}\right)\right)
\]
but, given convexity of $\Pi^{FI}_{t+1,l} (\cdot)$, we have

\[
\frac{\partial_{h}\Pi_{t+1}\left(N^{V}\left(\phi_{0}\right)\right)}{\pi_{hh}}
>
\frac{\partial_{l}\Pi_{t+1}\left(N^{V}\left(\phi_{0}\right)\right)}{\pi_{hl}},
\]
which contradicts (\ref*{eq:Diff_N_compare}).

Step 7 (item iv).
For any $\phi \in \Phi$, equation (\ref*{FOC_N}) implies
\[
    \left[ 
        \begin{matrix}
            \lambda \ell(\phi) \pi_{ll} - \mu \pi_{hl} \\         
            \lambda \ell(\phi) \pi_{lh} - \mu \pi_{hh}         
        \end{matrix}
     \right]
    \in 
    \partial \Pi_{t+1}(N^V ( \phi (y)))
\]
Considering any two $\phi,\phi' \in \Phi$, convexity of $\Pi_{t+1} (\cdot)$ implies:
\begin{align*}
        \Pi_{t+1}(N^V ( \phi' ))  -  \Pi_{t+1}(N^V ( \phi))
        \leq
        \left[ 
            \begin{matrix}
                \lambda \ell(\phi) \pi_{ll} - \mu \pi_{hl} \\         
                \lambda \ell(\phi) \pi_{lh} - \mu \pi_{hh}         
            \end{matrix}
         \right]^T
         [N^V (\phi') - N^V(\phi)] \\
        \Pi_{t+1}(N^V ( \phi ))  -  \Pi_{t+1}(N^V ( \phi'))
        \leq
        \left[ 
            \begin{matrix}
                \lambda \ell(\phi') \pi_{ll} - \mu \pi_{hl} \\         
                \lambda \ell(\phi') \pi_{lh} - \mu \pi_{hh}         
            \end{matrix}
         \right]^T
         [N^V (\phi) - N^V(\phi')]
\end{align*}

Summing up both equations gives us:
\[
    0 
    \leq 
    \lambda [\ell(\phi) - \ell(\phi')]
    \left[ 
        \begin{matrix}
            \pi_{ll} \\
            \pi_{lh}
        \end{matrix}
     \right]^T
    [N^V (\phi') - N^V(\phi)],
\]
which implies (iv).

\end{proof}

\section{Proof of Lemma \ref{Lem:Properties_relaxed_problem}}

\begin{proof}
    Uniqueness follows from Lemma \ref{lem:uniqueness-concavity}.

    The remaining statements are proved in the order: 
    $ii \rightarrow iv \rightarrow i \rightarrow iii \rightarrow v$.

    Statement (ii) follows directly from Lemma \ref{lem:show-recursivity}-(i), which has $\Pi_1=\Pi^*$ as a special case.

    All other properties are derived from Lemma \ref{lem:recursive-properties}.
    For any $t=1,\dots,T$ and $\eta^{t-1} = (h^{t-1}, \phi^{t-1})$, we have that:
    \begin{gather*}
        z_t(y \mid \eta^{t-1}, h) = \vartheta(y), \\
        V_{t+1}(\eta^{t-1},\phi_t,h,i) = N_i(\phi_t) \text{, for }i=l,h,
    \end{gather*}
    where $(\vartheta,N)$ is the solution to problem 
    \begin{equation} \label{eq:update-continuation-utility}
        P_t \left( 
            V_t(\eta^{t-1},l),
            V_t(\eta^{t-1},h)
        \right).
    \end{equation}
    
    Proof of (iv). Follows from $V_l<V_h$ and Lemma \ref{lem:recursive-properties}-(iii).

    Proof of (i). For any history $\eta^{t-1} = (h^{t-1},\phi^{t-1})$, the result follows from (iv), which states that the continuation utility following $\theta_t=h$ is strictly higher than that following $\theta_t=l$, and Lemma \ref{lem:recursive-properties}-(i), which implies that the within-period $t$ upward incentive constraint binds as a result.

    Let $\mu_{t-1}(\phi^{t-1})$ and $\lambda_{t-1}(\phi^{t-1})$ be the Lagrange multipliers of the problem (\ref{eq:update-continuation-utility}), with $\eta^{t-1} = (h^{t-1}, \phi^{t-1})$.

    Proof of (iii). Follows from Lemma \ref{lem:recursive-properties}-(ii), using the following relationship to the solution $\vartheta(\cdot)$ of problem (\ref{eq:update-continuation-utility}):
    \begin{equation*}
        \psi'(\vartheta(y_t)) = \frac{
            1
        }{
            u' \left( 
                z_t(y_t \mid \eta^{t-1},h)
            \right)
        }.
    \end{equation*}
    
    Proof of (v). Follows directly from Lemma \ref{lem:recursive-properties}-(iv).
\end{proof}

\section{Results on the auxiliary problem}

For simplicity, I write the auxiliary problem in terms of utility levels:
\begin{eqnarray*}
\chi\left(\nu,\Delta\right) & = & \inf_{x:Y\rightarrow u\left(\mathbb{R}_{+}\right)}\sum_{y}p_{h}\left(y\right)\psi\left[x\left(y\right)\right],
\end{eqnarray*}
subject to
\[
\sum_{y}p_{h}\left(y\right)x\left(y\right)=\nu,
\]
and
\[
\sum_{y}p_{l}\left(y\right)x\left(y\right)=\nu-\Delta.
\]

The following statement provides important properties of cost function $\chi$ that are used in the analysis.

\begin{lem}
    \label{Lem:Appendix_Chi_Properties}
    The problem $\mathcal{P}^A$ satisfies the following: \\
    (i) It has a unique solution, \\
    (ii) $\chi(\cdot)$ is strictly convex,\\
    moreover, if 
    $x\left(\cdot\right) 
    \in int\left[u\left(\mathbb{R}_{+}\right)\right]^{Y}$
    solves $\mathcal{P}^A$, then: \\
    (iii) $\chi\left(\nu,\Delta\right)$ is twice continuously differentiable
    in an open neighborhood of $\left(\nu,\Delta\right)$, \\
    (iv) $sign\left(\frac{\partial\chi\left(\nu,\Delta\right)}{\partial\Delta}\right)=sign\left(\Delta\right)$, \\
    (v) cross derivative sign: 
    \begin{equation*}
        sign\left(\frac{\partial^{2}\chi}{\partial \nu\partial\Delta}\right)
        =
        \begin{cases}
            sign\left(\Delta\right)\text{, if }\psi'''>0 \\
            -sign\left(\Delta\right)\text{, if }\psi''' < 0 \\
            = 0 \text{, if } \psi''' = 0,
        \end{cases}   
    \end{equation*}\\
    (vi)
    $    \psi'(x(y)) 
    = \chi_\nu (\nu, \Delta) 
    + \chi_\Delta (\nu, \Delta) \left[ 
        1 - \ell (y)
    \right] 
    $.
\end{lem}

\begin{proof}
Existence of solution follows from the fact that 
\[
\left\{ x\in\left[u\left(\mathbb{R}_{+}\right)\right]^{Y}\mid\sum_{y}p_{h}\left(y\right)\psi\left[x\left(y\right)\right]\leq K\right\} 
\]
is compact, for any $K\in\mathbb{R}_{+}$. 
Uniqueness and convexity (items i-ii) follows from the strict convexity of the objective function and linearity of the constraints in $(x,\nu,\Delta)$.

The following are necessary and sufficient conditions for 
$x\left(\cdot\right)\in int\left[u\left(\mathbb{R}_{+}\right)\right]^{Y}$
to be interior are: $\exists\lambda,\mu\in\mathbb{R}$ such that
\begin{equation}
    \psi'\left(x\left(y\right)\right)
    - \lambda
    + \mu \ell (y)
    =0,\label{eq:chiFOC1}
\end{equation}
\begin{equation}
\sum_{y\in Y}p_{h}\left(y\right)x\left(y\right)=\nu,\label{eq:chiFOC2}
\end{equation}
\begin{equation}
\sum_{y\in Y}p_{l}\left(y\right)x\left(y\right)=\nu-\Delta.\label{eq:chiFOC3}
\end{equation}
for all $y\in Y$. 

Consider $\left\{ y_{i}\right\} _{i\in I}$ an ordering of $Y$ such that $\left\{ \ell (y_i ) \right\} _{i\in I}$ is increasing. 
Then distributions 
$\left\{ p_{h}\left(y_{i}\right)\right\} _{i\in I}$
and $\left\{ p_{l}\left(y_{i}\right)\right\} _{i\in I}$ 
are ordered in terms of the monotone likelihood ratio property (MLRP). 
It then follows that $\left\{ x\left(y_{i}\right)\right\} _{i\in I}$ is decreasing (increasing) if $\mu>0$ ($\mu<0$), which implies that $\Delta$ must be strictly positive (negative). 
As a consequence, $sign\left(\mu\right)=sign\left(\Delta\right)$ (item iv). 

If $\left(\lambda,\mu,x\left(\cdot\right)\right)$ solve $\left(\ref{eq:chiFOC1}\right)-\left(\ref{eq:chiFOC3}\right)$,
then by the implicit function theorem the system has a unique continuously
differentiable solution $\left(\lambda^{\nu',\Delta'},\mu^{\nu',\Delta'},x\left(\cdot\mid \nu',\Delta'\right)\right)$
for $\left(\nu',\Delta'\right)$ in an open neighborhood of $\left(\nu,\Delta\right)$.
Therefore $\chi$ is continuously differentiable at $\left(\nu,\Delta\right)$,
and its derivative is given by

\begin{equation}
    \left[\begin{array}{c}
    \frac{\partial}{\partial \nu}\chi\left(\nu,\Delta\right)\\
    \frac{\partial}{\partial\Delta}\chi\left(\nu,\Delta\right)
    \end{array}\right]
    =\left[
    \begin{array}{c}
        \lambda^{\nu,\Delta}-\mu^{\nu,\Delta}\\
        \mu^{\nu,\Delta}
    \end{array}
    \right].
    \label{eq:EnvelopeChiFunction}
\end{equation}

Continuous differentiability of $\left(\lambda^{\nu,\Delta},\mu^{\nu,\Delta}\right)$
implies that $\chi\left(\cdot\right)$ is twice continuously differentiable
at $\left(\nu,\Delta\right)$ (item iii).

Finally, simple differentiation implies
\[
\frac{\partial^{2}\chi\left(\nu,\Delta\right)}{\partial \nu\partial\Delta}
=
\frac{\sum_{y}\frac{p_{l}\left(y\right)}{\psi''\left(x\left(y\right)\right)}-\sum_{y}\frac{p_{h}\left(y\right)}{\psi''\left(x\left(y\right)\right)}}{\left(\sum_{y}\frac{p_{h}\left(y\right)}{\psi''\left(x\left(y\right)\right)}\right)\left(\sum_{y}\frac{\left[p_{l}\left(y\right)\right]^{2}}{p_{h}\left(y\right)\psi''\left(x\left(y\right)\right)}\right)+\left(\sum_{y}\frac{p_{l}\left(y\right)}{\psi''\left(x\left(y\right)\right)}\right)^{2}}.
\]
Now assume $\psi'''>0$, 
$\left\{ \frac{1}{\psi''\left(x\left(y_{i}\right)\right)}\right\}_{i\in I}$
is increasing (decreasing) in $i$ if, only if, $\Delta>0$ ($\Delta<0$). 
Since 
$\left\{ p_{h}\left(y_{i}\right)\right\} _{i\in I}$
and $\left\{ p_{l}\left(y_{i}\right)\right\} _{i\in I}$ 
are MLRP ordered, we have that
\[
sign\left(\frac{\partial^{2}\chi\left(\nu,\Delta\right)}{\partial \nu\partial\Delta}\right)=sign\left(\Delta\right).
\]
An analogous argument proves the cases $\psi'''=0$ and $\psi'''<0$ (item v).
Item (vi) follows from (\ref{eq:chiFOC1}) and (\ref{eq:EnvelopeChiFunction}).
\end{proof}

\begin{lem} \label{lem:chi-derivatives-foc}
    The solution to $\mathcal{P}^A$ for the pair $(\nu, \Delta)$ satisfies 
    \begin{equation*}
        \frac{1}{
            u' (\zeta(y))
        } 
         = \chi_\nu (\nu, \Delta) 
        + \chi_\Delta (\nu, \Delta) \left[ 
            1 - \ell (y)
         \right].
    \end{equation*}
\end{lem}
\begin{proof}
    Follows from Lemma \ref{Lem:Appendix_Chi_Properties}, with the solution in terms of utility flows and consumption being connected via
    $
        \frac{1}{
            u' (\zeta(y))
        }
        = \psi'(x(y)).
    $
\end{proof}

The following statement connects variation in marginal cost of utility and distortions to the levels of utility and distortions and will be instrumental in the analysis of Subsection \ref{subsec:realization-independent-contracts}.

\begin{lem}
\label{lemma:single-crossing_chi}Suppose $\chi\left(\cdot\right)$
is strictly convex and $\frac{\partial^{2}\chi}{\partial \nu\partial\Delta}>0$,
then 
\[
\left\{ \begin{array}{c}
\chi_{\nu}\left(\nu,\Delta\right)\geq\chi_{\nu}\left(\nu',\Delta'\right),\\
\chi_{\Delta}\left(\nu,\Delta\right)\leq\chi_{\Delta}\left(\nu',\Delta'\right)
\end{array}\right\} \Rightarrow\left\{ \begin{array}{c}
\nu\geq \nu',\\
\Delta\leq\Delta'
\end{array}\right\} .
\]

Additionally, if the left conditions hold with strict inequalities,
then the implications also hold with strict inequalities.
\end{lem}
\begin{proof}
Since $\chi$ is twice continuously differentiable, the following holds:
\[
\chi_{\nu}\left(\nu,\Delta\right)-\chi_{\nu}\left(\nu',\Delta'\right)=\left(\nu-\nu'\right)\int_{0}^{1}\chi_{vv}\left(\iota\left(\alpha\right)\right)d\alpha+\left(\Delta-\Delta'\right)\int_{0}^{1}\chi_{\nu\Delta}\left(\iota\left(\alpha\right)\right)d\alpha\geq0,
\]
\[
\chi_{\Delta}\left(\nu,\Delta\right)-\chi_{\Delta}\left(\nu',\Delta'\right)=\left(\nu-\nu'\right)\int_{0}^{1}\chi_{\nu\Delta}\left(\iota\left(\alpha\right)\right)d\alpha+\left(\Delta-\Delta'\right)\int_{0}^{1}\chi_{\Delta\Delta}\left(\iota\left(\alpha\right)\right)d\alpha\leq0,
\]
where $\iota\left(\alpha\right)=\alpha\left(\nu,\Delta\right)+\left(1-\alpha\right)\left(\nu',\Delta'\right)$,
for $\alpha\in\left[0,1\right]$.

Using $ $$\frac{\partial^{2}\chi}{\partial \nu\partial\Delta}>0$, these imply
\[
\left(\Delta-\Delta'\right)\left[\frac{\int_{0}^{1}\chi_{\nu\Delta}\left(\iota\left(\alpha\right)\right)d\alpha}{\int_{0}^{1}\chi_{vv}\left(\iota\left(\alpha\right)\right)d\alpha}-\frac{\int_{0}^{1}\chi_{\Delta\Delta}\left(\iota\left(\alpha\right)\right)d\alpha}{\int_{0}^{1}\chi_{\nu\Delta}\left(\iota\left(\alpha\right)\right)d\alpha}\right]\geq0.
\]
However, convexity of $\chi\left(\cdot\right)$ second order continuous
differentiability, implies that the function $\Gamma$ defined over
a neighborhood of $\left(0,0\right)$, given by $\left(v_{0},\Delta_{0}\right)\mapsto\int_{0}^{1}\chi\left(f\left(\alpha\right)+\left(v_{0},\Delta_{0}\right)\right)d\alpha$
is also convex and twice continuously differentiable. Convexity implies
that 
\[
\left|\Gamma''\left(0,0\right)\right|=\left(\int_{0}^{1}\chi_{vv}\left(\iota\left(\alpha\right)\right)d\alpha\right)\left(\int_{0}^{1}\chi_{\Delta\Delta}\left(\iota\left(\alpha\right)\right)d\alpha\right)-\left(\int_{0}^{1}\chi_{\nu\Delta}\left(\iota\left(\alpha\right)\right)d\alpha\right)^{2}>0.
\]
This implies that
$\Delta \leq \Delta'$,
which together with $\chi_\nu (\nu, \Delta) \geq \chi_\nu (\nu',\Delta')$ implies $\nu \geq \nu'$.
Moreover, if the left inequalities in the Lemma hold strictly, we have similarly that $\Delta<\Delta'$ and $\nu > \nu'$.
\end{proof}

\subsection*{Proof of Lemma \ref{prop:Sufficient_supermodularity}} \label{proof:conditions-supermodularity}
\begin{proof}
    The equivalence of (i) and (iii) follows from Lemma \ref{Lem:Appendix_Chi_Properties}-(v).

    The second order derivative of $\psi=u^{-1}$ is given by
    \begin{eqnarray*}
    \psi''\left(x\right) & = & \left(-\frac{u''\left(\psi\left(x\right)\right)}{u'\left(\psi\left(x\right)\right)}\right)\frac{1}{u'\left(\psi\left(x\right)\right)^{2}}.
    \end{eqnarray*}

    The absolute risk aversion of utility $u\left(\cdot\right)$ at $c\in\mathbb{R}_{+}$
    is given by 
    \(
    r_u\left(c\right)\equiv-\frac{u''\left(c\right)}{u'\left(c\right)}.
    \)
    Therefore direct derivation implies that
    \[
    \psi'''\left(x\right)=\frac{r_u'\left(x\right)u'\left(x\right)+2\left[r_u\left(x\right)\right]^{2}}{\left[u'\left(x\right)\right]^{5}},
    \]
    where $x=u\left(x\right)$. This implies the equivalence between (ii) and (iii).
\end{proof}

\section{Utility and distortion dynamics}

In this section, we focus throughout on a particular $t=1,\dots,T-1$ and $\phi^{t-1} \in \Phi^{t-1}$ and study the problem of reallocating both flow utilities and distortions across periods $t$ and $t+1$, following series of announcements $\hat{\theta}^t=h^t$ and signals $\phi^{t-1}$.
For notational brevity, we will omit the dependence of distortions and flow utilities on $\phi^{t-1}$, denoting $\nu_t(\phi^{t-1})$ and $\nu_{t+1}(\phi^{t-1},\phi)$ simply as $\nu_t$ and $\nu_{t+1}(\phi)$, for example.

Define problem $\mathcal{P}^I$

\begin{equation*}
    \min_{(\nu,\Delta,(\nu'(\phi),\Delta'(\phi),\nu^l(\phi))_{\phi \in \Phi}) \in \bar{A}} 
    \chi (\nu, \Delta) + \delta \sum_{\phi \in \Phi} p_h(\phi) \left[ 
        \pi_{hh} \chi \left( \nu'(\phi), \Delta'(\phi) \right)
        + \pi_{hl } \chi(\nu^l(\phi), 0)
    \right]
\end{equation*}
subject to:

\begin{equation} \label{eq:Intertemporal_promise_keep}
    \nu + \delta \sum_{\phi \in \Phi} p_h(\phi) \left[ 
        \pi_{hh} \nu'(\phi)
        + \pi_{hl } \nu^l(\phi)
    \right]
    = V_t(h^{t-1},h),
\end{equation}

\begin{equation} \label{eq:Intertemporal_IC_t}
    \nu -\Delta
    + \delta \sum_{\phi \in \Phi} p_l(\phi) \left[ 
        \pi_{lh} \nu'(\phi)
        + \pi_{ll } \nu^l(\phi)
    \right]
    = V_t(h^{t-1},l),
\end{equation}
and 
\begin{multline} \label{Intertemporal_iC_t1}
    \nu'(\phi) -\Delta'(\phi)
    + \delta \sum_{\phi' \in \Phi} p_l(\phi') \left[ 
        \pi_{lh} V_{t+2}(h^{t+1},(\phi,\phi'),h)
        + \pi_{ll } V_{t+2}(h^{t+1},(\phi,\phi'),l)
    \right]
    = \\
    \nu^l(\phi)
    + \sum_{t' = t+2}^T \delta^{t'-t-1} u(c_{t+1}(\phi)).
\end{multline}

With the set $\bar{A}$ defined as 
\begin{equation*}
    \bar{A} \equiv \left\{ 
        (\nu,\Delta,(\nu'(\phi),\Delta'(\phi),\nu^l(\phi))_{\phi \in \Phi}) \mid 
        (\nu, \Delta),
        (\nu'(\phi), \Delta'(\phi)),
        (\nu^l(\phi), 0) \in A,
        \forall\phi\in \Phi
    \right\}.
\end{equation*}

\begin{lem}
    The vector
    \begin{equation*}
        \left( 
           \nu_t , \Delta_t,
           \left( 
               \nu_{t+1} (\phi) , \Delta_{t+1}(\phi),
               \nu^l_{t+1}(\phi)
            \right)_{\phi \in \Phi}
        \right),
    \end{equation*}
\end{lem}

solves problem $\mathcal{P}^I$.

\begin{proof}
    First notice that, from Proposition \ref{prop:Optimal_solves_Chi}, vector $\left( 
        \nu_t , \Delta_t,
        \left( 
            \nu_{t+1} (\phi) , \Delta_{t+1}(\phi),
            \nu^l_{t+1}(\phi)
         \right)_{\phi \in \Phi}
     \right)$ 
    is feasible in $\mathcal{P}^I$.

    For any $(\nu,\Delta,(\nu'(\phi),\Delta'(\phi),\nu^l)_{\phi \in \Phi}) \in \bar{A}$, consider a new mechanism that is identical to optimal mechanism $M$ except for changing:\\
    (i) $z_t(h^{t-1},h)$ to $\zeta(\nu,\Delta)$,\\
    (ii) $z_{t+1}(h^{t},\phi,h)$ to $\zeta(\nu'(\phi),\Delta'(\phi))$,\\
    (iii) consumption in $t+1$ following signals $(\phi^{t-1},\phi)$ and announcements $(h^t,l)$ to $\psi(\nu'(\phi))$.

    All incentive constraints in relaxed problem for announcements in $t' \geq t+2$ are unaffected.
    All incentive constraints in relaxed problem for announcements in $t' \leq t-1$ are unaffected since (\ref{eq:Intertemporal_promise_keep}) guarantees that the continuation payoff of the consumer at the start of period $t$ is unchanged.

    The period $t$ incentive constraint, following $\phi^{t-1}$, is guaranteed by (\ref{eq:Intertemporal_IC_t}), while all period $t+1$ incentive constraints are guaranteed by (\ref{Intertemporal_iC_t1}).

    Finally, the mechanism modification only affects the firm's expected discounted profits via the expression in the objective function of problem $\mathcal{P}^I$. Optimality of $M$ implies the result.

\end{proof}

\begin{proof}[Proof of Proposition \ref{prop:intertemporal_condition_general}]
    Using equations (\ref{eq:Intertemporal_promise_keep})-(\ref{Intertemporal_iC_t1}), we can simplify problem $\mathcal{P}^I$ to one where only flow utilities in period $t+1$ are chosen by finding expressions for distortions $(\Delta,(\Delta'(\phi))_{\phi \in \Phi})$ and period $t$ flow utility $\nu$, in terms of $(\nu'(\phi))_{\phi \in \Phi}$.

    By assumption, the solution of this problem is interior and hence the following local optimality conditions must be satisfied:

    \begin{equation} \label{Chi_FOC_1}
         \chi_v^1 
         + \chi_\Delta^1 \left[ 
            1 - \frac{\pi_{lh}}{\pi_{hh}} \ell(\phi)
         \right]
         =
         \chi_v^2(\phi) + \chi_\Delta^2(\phi),
    \end{equation}
    and 
    \begin{equation} \label{Chi_FOC_2}
        \chi_v^1 
        + \chi_\Delta^1 \left[ 
            1 - \frac{\pi_{ll}}{\pi_{hl}} \ell(\phi)
        \right]
        =
        \chi_v^{2,l}(\phi) 
        - 
        \frac{\pi_{hh}}{\pi_{hl}} \chi_\Delta^2 (\phi),
    \end{equation}
where $\chi_k^1 \equiv \chi_k(\nu_t,\Delta_t)$, 
$\chi_k^2(\phi) \equiv \chi_k(\nu_{t+1}(\phi),\Delta_{t+1}(\phi))$,
for $k\in \left\{ \nu,\Delta \right\}$, 
and 
$\chi_v^{2,l} \equiv \chi_v(\nu^l_{t+1}(\phi))$.

Multiplying (\ref{Chi_FOC_1}) by $\pi_{hh}p_h(\phi)$ 
and (\ref{Chi_FOC_2}) by $\pi_{hl} p_h(\phi)$,
adding both equations,
and finally 
summing across signals $\phi \in \Phi$ 
gives us (\ref{eq:Intertemporal_util_Chi_general}).
Subtracting (\ref{Chi_FOC_1}) from \ref{Chi_FOC_2}, multiplying the result by $p_h(\phi)$ and summing over $\phi \in \Phi$ gives us (\ref{eq:Intertemporal_delta_Chi_general}).
\end{proof}

\subsection{Realization-independent mechanisms}

For brevity, we now define  $\chi_k^t \equiv \chi_k(\nu_t,\Delta_t)$ and 
$\chi^{t,l}_\nu \equiv \chi_\nu(\nu^l_t,0)$ for 
for $k\in \left\{ \nu,\Delta \right\}$ and
$t=1,\dots,T$. We also introduce notation 
\begin{equation*}
    V^d_t \equiv V_t(h^{t-1},h) - V_t(h^{t-1},l),
\end{equation*}
which, using the binding period $t$ incentive constraint, satisfies
\begin{equation} \label{eq:Delta_cont_util}
    V^d_t = \Delta_t + \delta (\pi_{hh} - \pi_{lh}) V^d_{t+1},
\end{equation}
with $V^d_{T+1}=0$, if $T<\infty$.

\begin{lem} \label{lem:Condition_monotonicity_dynamics}
    For any $t=2,\dots,T$, if $V_h > V_l$, the optimal mechanism is interior and satisfies 
    $\chi^t_\nu > \chi_\nu^{t,l}$,
    then it must satisfy:
    \begin{equation*}
        \nu_{t-1} < \nu_t,
    \end{equation*}
    and
    \begin{equation*}
        \Delta_{t-1} > \Delta_t.
    \end{equation*}
\end{lem}

\begin{proof}
    The intertemporal optimality condition 
    \ref{eq:Intertemporal_util_Chi_RI}
    implies that
    \begin{equation*}
        \chi^t_\nu > \chi_\nu^{t-1} > \chi_\nu^{t,l},
    \end{equation*}
    while, using 
    $\chi^t_\nu > \chi_\nu^{t,l}$ and optimality condition 
    (\ref{eq:Intertemporal_delta_Chi_RI}) gives us
    \begin{equation*}
        \chi_\Delta^{t-1} < \chi_\Delta^{t}.
    \end{equation*}
    Hence, we have that 
    \begin{align*}
        \chi_\nu(\nu_t, \Delta_t) > \chi_\nu(\nu_{t-1}, \Delta_{t-1}),\\
        \chi_\Delta(\nu_t, \Delta_t) < \chi_\Delta(\nu_{t-1}, \Delta_{t-1})
    \end{align*}
    which, using Lemma \ref{lemma:single-crossing_chi}, imply the result.
\end{proof}

\begin{lem} \label{lem:Recursive_Chi}
    For any $t=2,\dots,T$, if $V_h > V_l$ then an interior optimal mechanism  satisfies: \\
    (i) $\chi^t_\nu > \chi_\nu^{t,l}$,\\
    (ii) $V^d_{t-1} > V^d_t$.
\end{lem}

\begin{proof}
    We prove this statement by induction.
    
    First, consider $t=T$. 
    In this case we have that
    \begin{equation*}
        \Delta_T = \nu_T - \nu^l_T,
    \end{equation*}
    and, since Proposition \ref{prop:Optimal_solves_Chi} guarantees that $\Delta_T>0$, supermodularity implies (i) as:
    \begin{equation*}
        \chi_\nu(\nu_T,\Delta_T) \geq  \chi_\nu(\nu_T,0) > \chi_\nu(\nu^l_T).
    \end{equation*}
    Lemma \ref{lem:Condition_monotonicity_dynamics} then implies that $\Delta_{T-1}> \Delta_T$.
    
    Property (ii) follows since $V^d_T=\Delta_t$ and $V^d_{T-1} = \Delta_{T-1} + \delta (\pi_{hh} - \pi_{lh})\Delta_T$.

    Now suppose that properties (i)-(ii) hold for all $t'=t+1,\dots,T$.

    We start by showing that property (i) holds.
    The continuation utility of a high-type consumer satisfies
    \begin{equation} \label{eq:Cont_util_high}
        V_t(h^{t-1},h) = \nu_t + \delta \left[ 
            \pi_{hh} V_{t+1}(h^t,h) + \pi_{hl} V_{t+1} (h^t,l) 
        \right],
    \end{equation}
    while the continuation utility of a lot-type consumer satisfies 
    \begin{equation} \label{eq:Cont_util_low}
        V_t(h^{t-1},l) = \sum_{\tau=t}^T \delta^{\tau - t} \nu^l_t.
    \end{equation}

    Substituting equations (\ref{eq:Cont_util_high}) and (\ref{eq:Cont_util_low}) into (\ref{eq:Delta_cont_util}) gives us the following, after some manipulation:

    \begin{equation} \label{eq:Induction_manipulate}
        \Delta_t = (\nu_t - \nu^l_t) 
        + \sum_{\tau>t}\delta^{\tau - t} (\nu^l_{t+1} - \nu^l_t)
        +\delta\pi_{lh}V_{t+1}^d
    \end{equation}

    Suppose, by way of contradiction, that 
    \begin{equation*}
        \chi^t_\nu < \chi_\nu^{t,l}.
    \end{equation*}
    But, given supermodularity of $\chi$, this inequality requires
    \begin{equation} \label{eq:Order_nu_contradiction}
        \nu^l_t > \nu_t.
    \end{equation}

    Also, from our inductive hypothesis, we know that $\chi^{t+1}_\nu > \chi^{t+1,l}_\nu$, which together with (\ref{eq:Intertemporal_util_Chi_RI}) implies that 
    \begin{equation} \label{eq:Order_nu_contradiction2}
        \chi^{t,l}_\nu > \chi^{t}_\nu 
        = \pi_{hh} \chi^{t+1}_\nu + \pi_{hl} \chi^{t+1,l}_\nu 
        > \chi^{t+1,l}_\nu
        \implies 
        \nu^l_t > \nu^l_{t+1}.
    \end{equation}
    Combining (\ref{eq:Induction_manipulate}), (\ref{eq:Order_nu_contradiction}) and (\ref{eq:Order_nu_contradiction2}) we have 

    \begin{equation*}
        \Delta_t < 
        \delta\pi_{lh} V^d_{t+1},
    \end{equation*}
    and, using (\ref{eq:Delta_cont_util}) once again, we have that 
    \begin{equation*}
        V^d_t \leq \delta \pi_{hh}V^d_{t+1},
    \end{equation*}
    which contradicts property (ii), which holds at period $t+1$ by our inductive assumption.

    We now prove property (ii).
    Since property (i) holds for any $t' \geq t$,  Lemma \ref{lem:Condition_monotonicity_dynamics} implies that $\Delta_{t'-1}>\Delta_{t'}$ for all $t \leq t' \leq T$.
    Now notice that 
    \begin{equation*}
        V^d_t = \sum_{\tau=t}^T 
        \left[ 
            \delta
            (\pi_{hh} - \pi_{lh})
         \right]^{\tau - t} 
        \Delta_\tau,
    \end{equation*}
    and hence we have 
    \begin{equation*}
        V^d_{t-1} - V^d_t = \sum_{s=0}^{T-t} 
        \left[ 
            \delta
            (\pi_{hh} - \pi_{lh})
         \right]^{s} 
         \left( 
             \Delta_{t+s-1}
             - \Delta_{t+s}
          \right)
          +
          \left[ 
            \delta
            (\pi_{hh} - \pi_{lh})
         \right]^{T-t} 
         \Delta_{T},
    \end{equation*}
    which is strictly positive since the series $\left\{ \Delta_\tau \right\}_{\tau=t-1}^T$ is strictly increasing.
\end{proof}

\begin{proof}[Proof of Proposition \ref{prop:RI_monotonicity}] \label{Proof_prop_4}
    \hfill \\ 
    Follows directly from Lemmas \ref{lem:Condition_monotonicity_dynamics} and \ref{lem:Recursive_Chi}
\end{proof}



\section{Competitive analysis}

The two firms are labeled $A$ and $B$.
Let $V^E=(V^E_l, V^E_h)$ denote the equilibrium utility level of the consumer conditional on her initial type, and $V^j=(V^j_l, V^j_h)$ denote the vector describing the utility level obtained the consumer with both possible initial types when accepting the equilibrium offer of firm $j = A,B$. Finally, we denote the equilibrium profit of firm $j=A,B$ as $\Pi^*_j$.

\begin{lem} \label{lem:RS_unique}
    Any pure strategy PBE has outcome $M^{V^*}$.   
\end{lem}

\begin{proof}
    
    The proof is divided into four parts.

    I) $\Pi(V^E) = 0$ and both firms make zero profits in equilibrium.

    Optimality of the consumer's acceptance strategy implies that, for $j=A,B$ and $i=l,h$, the following hold:\\
    - $V^j_i \leq V^E_i$, \\
    - $V^j_i = V^E_i$ if $j$'s offer is accepted with positive probability by type $i$.

    Hence, if firm $j$'s offer is accepted by consumer with type $i$ with positive probability, its continuation profits are at most 
    \begin{equation*}
        \Pi_i(V^E_i,V^j_{i'}) \leq \Pi_i(V^E_i,V^E_{i'}),
    \end{equation*}
    as increasing the utility of type $i' \neq i$ increases the profit opportunities from the consumer with type $i$.
    This means that the total profits obtained by firms is at most
    \begin{equation*}
        \sum_{i=l,h}\pi_i\Pi_i(V^E) = \Pi(V^E).
    \end{equation*}
    However, by offering mechanism $M^{V'}$ with $V' = V^E +(\epsilon, \epsilon)$, for $\epsilon>0$ sufficiently small, each firm can guarantee profits $\Pi(V')$.
    In equilibrium, the offer made by each firm must dominate $M^{V'}$.
    Since profit function $\Pi$ is continuous, we have 
    $\lim_{\epsilon \rightarrow 0} \Pi(V^E + (\epsilon, \epsilon)) = \Pi(V^E)$.
    Combining the two implications, we have
    \begin{equation*}
        \Pi(V^E)
        \geq
        \Pi^*_A + \Pi^*_B
        \geq
        2\Pi(V^E).
    \end{equation*}
    This implies that $\Pi(V^E) = 0$ and that both firms make zero profits.

    \bigskip

    II) $V^E_i \geq V^{FI}_l$, for $i=l,h$.

    First suppose that, by way of contradiction, $V^E_l < V^{FI}_l$. In this case firm $A$ could guarantee positive profits by offering a mechanism with non-contingent constant consumption satisfying 
    \begin{equation*}
        u(\bar c)\sum_{{t=1}}^T\delta^{t-1}
        =
        V^E_l +\epsilon,
    \end{equation*}
    for $\epsilon>0$ sufficiently small, since it makes positive profits from the $l$-type consumer and, if the $h$-type consumer were to choose this contract, it would also generate positive profits since the discounted average income of the $h$-type consumer is strictly higher than that of the $l$-type consumer. 

    Finally, if $V^E_h < V^{FI}_l$, a similar profitable deviation exists.

    \bigskip

    III) $\Pi_h(V^E) \leq 0$.

    Suppose, by way of contradiction, that 
    $\Pi_h(V^E) > 0$.
    This implies that $V^E_h < V^{FI}_h$ and hence concavity of the feasible utility set $\mathcal{V}$ implies that, for $\epsilon >0$ sufficiently small, $(V^E_l,V^E_h + \epsilon) \in \mathcal{V}$.

    Also, we know that $V^E_l \geq V^{FI}_l>\sum_{t=1}^T \delta^{t-1} u(0)$ and hence incentive compatibility implies that both types' equilibrium utility are strictly above $\sum_{t=1}^T \delta^{t-1} u(0)$. 
    As a consequence, an incentive compatible mechanism in which the utility of both consumers is reduced exists, i.e., there exists $V' \in \mathcal{V}$ such that $V'_i < V^E_i$, for $i=l,h$.

    The existence of feasible utility vectors $(V^E_l,V^E_h + \epsilon)$, for $\epsilon>0$ and $V'$, together with convexity of $\mathcal{V}$, implies that: for $\varepsilon>0$ sufficiently small, there exists utility vector $\tilde{V} \in \mathcal{V}$ such that 
    \begin{gather*}
        V^E_l - \varepsilon < \tilde{V}_l < V^E_l,\\
        V^E_h + \varepsilon > \tilde{V}_h > V^E_h.
    \end{gather*}
    Considering $\varepsilon$ sufficiently small, a deviation to offer $M^{\tilde{V}}$ leads to profits approximately equal to
    \begin{equation*}
        \mu_h \Pi_h(V^E)>0,
    \end{equation*}
    which contradicts point I.

    \bigskip
    IV) Points II and III imply that total profits are non-positive, conditional on any realization of the consumer's initial type. Combined with I, it implies that
    \begin{equation*}
        \Pi_i(V^E) = 0,
    \end{equation*}
    for both $i=l,h$.
    Using II, we must have $V^E_l = V^{FI}_l$. Finally, since $\Pi_h(V^E_l , \cdot)$ is strictly decreasing in $(V^{FI}_l,\infty) \cap \mathcal{V}$, the unique possible utility $V^E_h$ is $V^*_h$.

\end{proof}

\begin{lem} \label{lem:RS_existence}
    A pure strategy equilibrium exists if, and only if,
    \begin{equation*}
        {u'\left[
            u^{-1}\left(
                c^{FI}_l
            \right)
        \right]}
        {
            \frac{\partial_{+}\Pi_{h}\left(V^{*}\right)}{\partial V_{l}}
        }
        <
        \frac{\mu_{l}}{\mu_{h}}.
    \end{equation*}
\end{lem}

\begin{proof}
    (If) Consider the following strategy profile. 
    Both firms offer mechanism $M^{V^*}$ and consumers follow an optimal equilibrium that follows truth-telling whenever optimal and treats firms symmetrically.

    Given the definition of $V^*$, no firm has a profitable deviation in which a single type is attracted to its offer. This is trivially the case for type $l$. For type $h$, such a deviation would require that it offer utility pair $V'$ with $V'_h>V^*_h$ and $V'_l > V^*_l$, which implies $\Pi_h(V')<\Pi_h(V^E) = 0$.

    Hence, this strategy profile constitutes a pure strategy equilibrium if, and only if, no alternative mechanism can attract both types and generate strictly positive expected profits.
    But given concavity of $\Pi(\cdot)$, we can find $V' \geq V^*$ such that 
    \begin{equation*}
        \Pi(V') > \Pi(V^*).
    \end{equation*}
    if, and only if, we can find a pair $(d_h,d_l)\geq 0$ such that
    \begin{equation*}
        \frac{\partial \Pi}{\partial V_l} (V^*) d_l
        +
        \frac{\partial_+ \Pi}{\partial V_h} (V^*) d_h
        >0,
    \end{equation*}
    where we have used the fact that $\Pi(\cdot )$ is differentiable in $V_l$, whenever $V_l<V_h$.
    Since 
    $\frac{\partial_+ \Pi}{\partial V_h} (V^*) < 0$, such a pair exists if, and only if, 
    $\frac{\partial \Pi}{\partial V_l} (V^*)>0$.
    
    Condition 
    $\frac{\partial \Pi}{\partial V_l} (V^*)>0$
    coincides with the expression in the lemma since, for $V$ with $V_l<V_h$
    \begin{equation*}
        \Pi(V) = \mu_l \Pi^{FI}(V_l) + \mu_h \Pi_h(V_l,V_h).
    \end{equation*}
    (Only if) If the condition in the Lemma fails, we can find $V' \geq V^*$ such that $\Pi(V') > \Pi(V^*)$.
    Now consider any pure strategy equilibrium with outcome $M^{V^*}$. Offer $M^{V'}$ is a profitable deviation by any firm.
\end{proof}
\end{document}